\documentclass[alpha-refs]{wiley-article}

\usepackage{siunitx}

\usepackage{graphicx}  
\usepackage{dcolumn}   
\usepackage{bm}        
\usepackage{natbib}
\usepackage{adjustbox}
\usepackage{amsfonts}

\newcommand{\R}{\mathbb{R}}

\usepackage{amssymb,latexsym,amssymb,amsmath,amsbsy,amsopn,amstext,upgreek}
\usepackage{color}
\usepackage{graphicx,wrapfig,fancybox,watermark,graphics}
\usepackage{pgf}
\usepackage{float}

\usepackage{vmargin}
\usepackage{cjhebrew}
\usepackage{subfigure}
\usepackage[toc,page]{appendix}

\newcommand{\healpix}{HEALpix~}
\newcommand{\nside}[1]{N_{\rm side}}

\title{A Numerical Investigation on the High-Frequency Geometry of Spherical Random Eigenfunctions}

\author[1\authfn{1}]{Yabebal Fantaye}
\author[2\authfn{1}]{Valentina Cammarota}
\author[1\authfn{2}]{Domenico Marinucci}
\author[2 \authfn{2}]{Anna Paola Todino}

\affil[1\authfn{1}]{African Institute for Mathematical Sciences and Stellenbosch University, Cape Town, South Africa}
\affil[2\authfn{1}]{Department of Statistics, Sapienza University of Rome, Italy}
\affil[2\authfn{2}]{Gran Sasso Science Institute, L'Aquila, Italy}

\corraddress{1\authfn{2}Domenico Marinucci, Department of Mathematics, Tor Vergata, University of Rome, Via della Ricerca Scientifica 1, I-00133 Roma, Italy}
\corremail{marinucc@mat.uniroma2.it}

\fundinginfo{The authors acknowledge support from ERC Grant 277742 Pascal.  YF is supported
	by the Robert Bosch Stiftung. The research by DM was supported by the MIUR Excellence Department Project awarded to the Department of Mathematics, University of Rome Tor Vergata, CUP E83C18000100006. }

\begin{document}

\maketitle

\begin{abstract}
A lot of attention has been drawn over the last few years by the investigation of the geometry of spherical random eigenfunctions (random spherical harmonics) in the high frequency regime, i.e ., for diverging eigenvalues. In this paper, we present a review of these results and we collect for the first time a comprehensive numerical investigation, focussing on particular on the behaviour of Lipschitz-Killing curvatures/Minkowski functionals (i.e., the area, the boundary length and the Euler-Poincar\'e characteristic of excursion sets) and on critical points. We show in particular that very accurate analytic predictions exist for their expected values and variances, for the correlation among these functionals, and for the cancellation that occurs for some specific thresholds (the variances becoming an order of magnitude smaller - the so-called Berry's cancellation phenomenon). Most of these functionals can be used for important statistical applications, for instance in connection to the analysis of Cosmic Microwave Background data.
\end{abstract}

\section{Introduction}

The geometry of the excursion sets for random fields on the sphere (to be defined below) has been the object of rather intense research over the last decade or so. 
It is well-known that these geometrical properties can be characterized in terms of the so-called Lipschitz-Killing curvatures (or equivalently, Minkowski functionals), which in the two dimensional case correspond to the excursion area, (half) the boundary length and the Euler-Poincar\'e characteristic (connected regions minus number of holes) of the excursion set. A comprehensive description of Lipschitz-Killing curvatures for excursion sets of random fields is given in the excellent monograph by \cite{TaylorAdler2009}; these functionals can be computed on real data by means of
accurate and numerically efficient algorithms \cite{KLENK2006127, Guderlei2007, Gay2012_NGPeaks}. They have been widely applied in the analysis of experimental data, especially in a Cosmological framework, see i.e.,  \cite{natoli2010mfBoomrang, matsubara2010_mfFnl,
  ducout2013_mfFnl, gratten2012_mfLSSreview, munshi2013_mfSkewCl,planck2013_IS, planck2014-a18} and the references therein.

A lot of mathematical efforts have been spent since the '80s on the characterization of expected values of these functionals under Gaussianity, culminating in the discovery of the beautiful Gaussian Kinematic Formula (\cite{TaylorAdler2009}); comparing these expected values with realizations allows the implementation of a number of tests for Gaussianity and Isotropy (see again \cite{planck2013-p09, planck2014-a18}).

While the behaviour of expected values is now fully understood, it is clear that the implementation of more sophisticated, hence more sensitive, testing procedures requires further knowledge, in particular the variance of these functionals and therefore the possibility to establish Central Limit Theorems with correct normalization factors. Establishing a Central Limit result requires of course the exploitation of a suitable notion of asymptotic behaviour; in the framework of spherical fields, the only relevant notion seems to be the one of High-Frequency asymptotics. In particular, it is well-known that isotropic random fields on the sphere can be decomposed by means of the Spectral Representation Theorem into the sum of orthogonal components, each of them corresponding to a different multipole $\ell$. The behaviour of geometric functionals in the high-frequency/high energy limit for these components has been studied by several authors in recent years, starting from \cite{NS09,NS16} for the number of connected components, \cite{W09} for the nodal length, and then including, among others, \cite{MW11} and \cite{MR15} for the excursion area, \cite{MW14}, \cite{MW12} and \cite{R16} for the Defect, \cite{CM18} for the Euler-Poincar\'e characteristic, \cite{W09}, \cite{MRW17}, \cite{MPRW16}, \cite{R18} for the distribution of the nodal length, \cite{CMW16} for the critical values and \cite{CW17} for the total number of critical points (see also \cite{KKW13}, \cite{RW16}, \cite{DNPR16,M17a,RW18,PR18} for related works covering also the 2-dimensional torus, \cite{M17b} for the 3-dimensional torus and \cite{NPR17} for planar random waves).

Our aim in this paper is to present a unified overview of the literature, and especially to perform a detailed numerical investigation to verify the practical relevance of these results when investigating spherical Gaussian maps. We shall address several issues concerning not only the expected value and variances of Minkowski functionals, but also their cross-correlation across different level sets. The theoretical predictions which have so far been produced are validated for the first time from a numerical point of view, and moreover their domain of applicability is clarified. Indeed, in terms of the variances the theoretical expressions which are obtained should be viewed as leading terms in series expansions of the variances over different ``chaos" components; as such, the approximation depends on the rate of convergence to zeroes of the terms which are dropped. These rates are known to be polynomial in some cases (namely, those corresponding to non-zero levels) and logarithmic in others (those corresponding to zero levels); this duality is mirrored in the numerics that we shall present below.

Minkowski functionals are not the only objects of interest in this paper. Indeed, some other recent contributions have derived neat analytic formulae for the expected number and the variance of critical points on the same spherical harmonics components as considered earlier for Minkowski functionals. We are hence providing for the first time numerical evidence also on these statistics.

The plan of the paper is as follows: in Section \ref{second sec}, we review the results on the expected values and variances for the Lipschitz-Killing curvatures; in Section \ref{third sec} we discuss the behaviour of critical points, again reviewing the analytic results that are currently available, while Section \ref{corr sec} is devoted to the analysis of the correlation among these different functionals. Section \ref{sec:numerical} describes our implementation algorithms and presents the numerical results; we then draw some conclusions and present directions for future work.

\section{Characterization of Excursion Sets for Random Spherical Harmonics}\label{second sec}

In the case of the two-dimensional sphere, the excursion sets $A_{u}(f)$ of a given (possibly random)
function $f$ are defined for any real number $u$ as
\begin{equation}
A_{u}(f):=\left\{ x\in \mathbb{S}^{2}:f(x)\geq u\right\} \text{ .}
\end{equation}
Of course, in the limit where we take $u=-\infty ,$ we have that $%
A_{u}(f)=\mathbb{S}^{2}$. In this paper, we shall be concerned with random eigenfunctions $f_\ell$ which satisfy the Helmhotz equation

\begin{equation*}
\Delta _{\mathbb{S}^{2}}f_{\ell }+\lambda _{\ell }f_{\ell }=0,\mbox{ }\mbox{
}\mbox{ }f_{\ell }:\mathbb{S}^{2}\rightarrow \mathbb{R},
\end{equation*}%
where $\Delta _{\mathbb{S}^{2}}$ is the Laplace-Beltrami operator on $%
\mathbb{S}^{2}$, defined as usual as
\begin{equation*}
\dfrac{1}{\sin \theta }\dfrac{\partial }{\partial \theta }\bigg\{\sin \theta 
\dfrac{\partial }{\partial \theta }\bigg\}+\dfrac{1}{\sin ^{2}\theta }\dfrac{%
\partial ^{2}}{\partial \varphi ^{2}},\mbox{ }0\leq \theta \leq \pi ,\mbox{ }%
0\leq \varphi < 2\pi ,
\end{equation*}%
and $\lambda _{\ell }=\ell (\ell +1),\mbox{ }\ell =0,1,\dots $. For a given
eigenvalue $\lambda _{\ell },$ the corresponding eigenspace is the $(2\ell
+1)-$dimensional space of spherical harmonics of degree $\ell .$
The random fields $\{f_{\ell }(x),x\in \mathbb{S}^{2}\}
$ are Gaussian and isotropic with zero mean 
$ 
	\mathbb{E}[f_{\ell }(x)] =0 $ and variance $\mathbb{E}[f_{\ell }(x)^{2}]=1.$ The covariance function is given by $$
	\mathbb{E}[f_{\ell }(x)f_{\ell }(y)] =P_{\ell }(\cos d(x,y))$$
where $P_{\ell }$ are the Legendre polynomials and $d(x,y)$ is the spherical
geodesic distance between $x$ and $y,$ i.e. 
\begin{equation*}
d(x,y)=\arccos (\langle x,y\rangle ).
\end{equation*}%

Spherical random eigenfunctions are of interest because they can also be interpreted as the harmonics/Fourier components of data observed on the sphere. Indeed, let us first recall the well-known Spectral Representation Theorem for spherical random fields, which states that the following identity holds, in the $L^2$ sense:
\begin{equation}
f(x)=\sum_{\ell =0}^{\infty}\sum_{m=-\ell }^{\ell }a_{\ell m}Y_{\ell
m}(x)=\sum_{\ell =0}^{\infty}\sqrt{\frac{(2\ell+1)C_{\ell}}{4\pi}}f_{\ell }(x)\text{ ;}  \label{specrap}
\end{equation}%
here, the sequence $\{C_{\ell}\}$ denotes the so-called angular power spectrum of the field. The spherical harmonics coefficients may be computed from the field $f(.)$ by means of the inverse transform

\begin{equation}
a_{\ell m}= \int_{\mathbb{S}^2} f(x) \bar{Y}_{\ell
m}(x)dx, \text{ } \text{ } \ell =1,2,...., m=-\ell,...,\ell ,   \label{specrap2}
\end{equation}%
with $\mathbb{E}[a_{\ell m}]=0 $ and $\mathbb{E}|a_{\ell m}|^2=C_\ell$.  
The inverse transform \eqref{specrap2} is only feasible for unmasked (full-sky) data, a condition which is usually considered rather difficult to meet for astrophysical experiments such as those concerning Cosmic Microwave Background radiation. Rather recently, however, full-sky maps were produced for instance by \cite{starketal2014} and by the Planck collaboration in its 2018 release (see \cite{planck2018missionpaper}). 


Let us now recall again the definitions of the \emph{Lipschitz-Killing Curvatures} (LKCs), which correspond to Minkowski functionals up to a different indexing and normalization factors; in two dimension, they are given by  (a) the Euler-Poincar\'{e} characteristic (written $\mathcal{L}_{0}(A_{u}(f))$), e.g. the number of connected
regions minus the number of holes; (b) half the boundary length of
the excursion regions (written $\mathcal{L}_{1}(A_{u}(f))$); the area of the excursion
regions (written $\mathcal{L}_{2}(A_{u}(f))$), which corresponds to the first Minkowski functional. The expected values of these functionals when evaluated on the excursion sets of Gaussian fields have been fully characterized by the Gaussian Kinematic Formula (GKF), see \cite{TaylorAdler2009}. 

We now need the family of functions $\rho _{l}(u)$, for $ l \in \mathbb{N}$, defined as
\begin{equation}
\rho _{l}(u)=(2\pi )^{-(l+1)/2}H_{l-1}(u)e^{-u^{2}/2}\text{ ,}
\end{equation} where $H_{k}(u)$, $k \in \mathbb{N}$, denotes as usual the family of Hermite polynomials, that is,
\begin{equation}
H_{0}(u)=1, H_{1}(u)=u, H_{2}(u)=u^{2}-1, \dots;
\end{equation}
it is convenient to define also
\begin{equation}
H_{-1}(u)=\sqrt{2\pi }(1-\Phi (u))e^{u^{2}/2},
\end{equation}
where $\Phi (u)$ is the Gaussian cumulative distribution function, whence%
\begin{eqnarray}
\rho _{0}(u) &=&(2\pi )^{-1/2}\sqrt{2\pi }(1-\Phi
(u))e^{u^{2}/2}e^{-u^{2}/2}=1-\Phi (u) \\
\rho _{1}(u) &=&\frac{1}{2\pi }e^{-u^{2}/2}\text{ , }\rho _{2}(u)=\frac{1}{%
\sqrt{(2\pi )^{3}}}ue^{-u^{2}/2}.
\end{eqnarray}%
\cite{TaylorAdler2009} write these components as $\mathcal{M}_{k}([u,\infty ))=\frac{1}{\sqrt{2\pi }}%
H_{k}(u)e^{-u^{2}/2}$ and denote them Gaussian Minkowski functionals. The so-called ``flag" coefficients are instead given by
\begin{equation}
\left[
\begin{array}{c}
i+l \\
l%
\end{array}%
\right] =\left(
\begin{array}{c}
i+l \\
l%
\end{array}%
\right) \frac{\omega _{i+l}}{\omega _{i}\omega _{l}}\text{ , for }\omega
_{i}=\frac{\pi ^{i/2}}{\Gamma (\frac{i}{2}+1)}\text{ ,}
\end{equation}%
that is, $\omega _{i}$ represents the area of the $i-$dimensional unit ball, $%
\omega _{1}=2,$ $\omega _{2}=\pi ,$ $\omega _{3}=\frac{4}{3}\pi$ and $\Gamma(\cdot)$ being the Gamma function $\Gamma(n+1)=n\Gamma(n)$.
As a last ingredient, we write $\lambda$ for the parameter which represents the second derivative of
the covariance function at the origin.

We are now ready to present the general expression for the expected value of Lipschitz-Killing curvatures of a process $f$ on a manifold $D$, i.e., the Gaussian Kinematic Formula which reads (Theorem 13.2.1 in \cite{TaylorAdler2009}):

\begin{equation}
\lambda ^{i/2}\mathbb{E}\mathcal{L}_{i}(A_{u}(f(x);D))=\sum_{l=0}^{\dim
(D)-i}\left[
\begin{array}{c}
i+l \\
l%
\end{array}%
\right] \lambda ^{(i+l)/2}\rho _{l}(u)\mathcal{L}_{i+l}(D)\text{ .}
\label{GKF}
\end{equation}

As an application of the previous result, let us consider the Fourier components $\{f_\ell(\cdot)\}_{\ell=1,2,\dots}$ normalized to have variance one; the GKF yields immediately (compare
\cite{MV16}, Corollary 5, see also \cite{CX16})%

\begin{equation}
\mathbb{E}\mathcal{L}_{0}(A_{u}(f_{\ell }(.);\mathbb{S}^{2}))=2\left\{ 1-\Phi
(u)\right\} +\frac{\lambda_{\ell}}{2}\frac{ue^{-u^{2}/2}}{\sqrt{(2\pi )^{3}}}%
4\pi \text{ ;}  \label{sh1}
\end{equation}
\begin{equation}
\mathbb{E}\mathcal{L}_{1}(A_{u}(f_{\ell }(.);\mathbb{S}^{2})) =\frac{\pi }{2}%
\frac{1}{\sqrt{2}}\lambda_{\ell}^{1/2}\frac{e^{-u^{2}/2}}{2\pi }%
4\pi=\frac{\pi}{\sqrt{2}} \lambda_{\ell}^{1/2}e^{-u^{2}/2}\text{ ;} \label{sh2}
\end{equation}
and%
\begin{equation}
\mathbb{E}\mathcal{L}_{2}(A_{u}(f_{\ell }(.);\mathbb{S}^{2}))=4\pi \times \left\{
1-\Phi (u)\right\} \text{ .}  \label{sh3}
\end{equation}

Of course, in order to exploit Lipschitz-Killing curvatures/Minkowski functionals to implement data analysis tools the expected value by itself is not sufficient, but we need also analytic expression for the variance. The latter was derived in some recent results by \cite{CMW16, CM18}; see \cite{Todino} for a review.

For our purposes, the results in these papers can be summarized as follows; the asymptotic behaviour of each of the three Lipschitz-Killing curvatures, evaluated on the excursion sets of random spherical harmonics, is dominated by a single, fully degenerate component, which can be written as:

\begin{equation*}
\mathtt{Proj}[\mathcal{L}_{k}(A_{u}(f_{\ell };\mathbb{S}^{2}))|2]
\end{equation*}%
\begin{equation}
=\frac{1}{2}\left[ 
\begin{array}{c}
2 \\ 
k%
\end{array}%
\right] \left\{ \frac{\lambda _{\ell }}{2}\right\}
^{(2-k)/2}H_{1}(u)H_{2-k}(u)\phi (u)\frac{1}{(2\pi )^{(2-k)/2}}\int_{\mathbb{%
S}^{2}}H_{2}(f_{\ell }(x))dx+a_{k}(\ell ),  \label{2GKF}
\end{equation}%
where
\begin{equation*}
a_{k}(\ell )=\left\{ 
\begin{array}{cc}
O_{p}(\ell ) & \text{for }k=0, \\ 
0 & \text{for }k=1,2%
\end{array}%
\right. \text{ .}
\end{equation*}%
Here, and in the sequel, we use $\mathtt{Proj}[.|q]$ for the projection of random quantities on the so-called Wiener chaoses of order $q$; the latter are spaces generated by linear combinations of Hermite polynomials of order $q$, computed in $f_{\ell}$ and its derivatives (we refer to \cite{NP11, MPRW16}, \cite{CM18} and the references therein for more discussions and details). It is also important to notice that $\frac{\lambda _{\ell }}{2}=P_{\ell }^{\prime
}(1)$ represents the derivative of the covariance function of random
spherical harmonics at the origin, so that the term%
\begin{equation*}
\frac{\lambda _{\ell }}{2}\int_{\mathbb{S}^{2}}H_{2}(f_{\ell }(x))dx
\end{equation*}%
can be viewed as a (random) measure of the sphere induced by the Riemannian
metric, somewhat in analogy with the interpretation given for the Gaussian Kinematic Formula on the expected value in the book by \cite{TaylorAdler2009}; recall indeed that for eigenfunctions $f_{\ell}$
on the sphere $\mathbb{S}^{2}$ the term ${\mathcal{L}}_{2}
(\mathbb{S}^{2})$ which appears in (\ref{GKF}) is exactly given by the area
of the sphere with radius $\left\{ \frac{\lambda _{\ell }}{2}\right\}
^{1/2}, $ i.e., 
\begin{equation*}
{\mathcal{L}}_{2}(\mathbb{S}^{2})=\frac{\lambda _{\ell }}{2}%
\times 4\pi =\frac{\lambda _{\ell }}{2}\int_{\mathbb{S}^{2}}H_{0}(f_{\ell
}(x))dx\text{ .}
\end{equation*}%
As was noted in \cite{CM18}, the Gaussian Kinematic Formula can be rewritten with a very similar expression to (\ref{2GKF}), i.e.: 
\begin{equation*}
\mathtt{Proj}[\mathcal{L}_{k}(A_{u}(f_{\ell };\mathbb{S}^{2}))|0]
\end{equation*}%
\begin{equation}
=\left[ 
\begin{array}{c}
2 \\ 
k%
\end{array}%
\right] \left\{ \frac{\lambda _{\ell }}{2}\right\} ^{(2-k)/2}H_{1-k}(u)\phi
(u)\frac{1}{(2\pi )^{(2-k)/2}}\int_{\mathbb{S}^{2}}H_{0}(f_{\ell
}(x))dx+b_{k}(\ell )\text{ ,}  \label{1GKF}
\end{equation}%
where 
\begin{equation*}
b_{k}(\ell )=\left\{ 
\begin{array}{cc}
2(1-\Phi (u))=O(1) & \text{for }k=0, \\ 
0 & \text{for }k=1,2%
\end{array}%
\right. .
\end{equation*}%
More explicitly (see also \cite{MW11, MW14, MR15}, \cite{Rossi}), we have the following analytic expressions for the leading term components of the LKCs (expected values and dominant stochastic term):

\vspace{0.5cm}

\emph{\bf  a) Excursion Area ($k=2$)} 
  As explained above, the expected value for the excursion area can be obtained (as for the other Lipschitz-Killing curvatures) by a simple application of the Gaussian Kinematic Formula, which yields:
\begin{eqnarray*}
\mathtt{Proj}[\mathcal{L}_{2}(A_{u}(f_{\ell };\mathbb{S}^{2}))|0]&=&\left\{ 
\frac{\lambda _{\ell }}{2}\right\} ^{0}\left[ H_{-1}(u)\phi (u)\right] \int_{%
\mathbb{S}^{2}}H_{0}(f_{\ell }(x))dx
\\
&=& 
\left[ 1-\Phi (u)\right] 4\pi \text{ ;}
\end{eqnarray*}%
the leading term in the fluctuations is provided by (see \cite{MW11, MR15})
\begin{equation*}
\mathtt{Proj}[\mathcal{L}_{2}(A_{u}(f_{\ell };\mathbb{S}^{2}))|2]=\frac{1}{2}%
\left\{ \frac{\lambda _{\ell }}{2}\right\} ^{0}\left[ H_{0}(u)H_{1}(u)\phi
(u)\right] \int_{\mathbb{S}^{2}}H_{2}(f_{\ell }(x))dx\text{ ;}
\end{equation*}
with an asymptotic variance which is given by 
\begin{eqnarray}\label{var nodale}
  \text{Var} \left( \mathcal{L}_{2}(A_{u}(f_{\ell };\mathbb{S}^{2})) \right) 
 &=& 16 \pi^2 \frac{u^2}{4}\phi^2(u)\frac{2}{2\ell+1} \\
 &=& 2\pi u^2e^{-u^2}\frac{1}{\ell} + o(\frac{1}{\ell}),
 \end{eqnarray}
where we have used the fact that (see \cite{MW11, MW14, Rossi}) 
\begin{equation}\label{Varh2}
     \text{Var} \left( \int_{\mathbb{S}^{2}}H_{2}(f_{\ell }(x))dx \right) = 16 \pi^2\frac{2}{2\ell + 1}.
 \end{equation}
Analogous results, although with different constants, can be established on subdomains of the sphere (see \cite{todino1}). For $u=0$, we obtain a quantity equivalent to the so-called defect  (see \cite{MW14}) i.e.,
$$D_\ell=2 \mathcal{L}_2(A_{u=0}(f_\ell; \mathbb{S}^2))-4\pi;$$
the expected value is immediately seen to be zero, while it can be shown that the variance is given by 
\begin{equation}\label{defect var}
 \text{Var}(D_\ell)=\dfrac{C}{\ell^2}+o\bigg(\dfrac{1}{\ell^2}\bigg),
\end{equation} 
where the constant $C$ can be computed as
\begin{equation}\label{C e a}
C=32\pi \sum_{k=1}^{\infty} a_k C_{2k+1} \mbox{ } \mbox{ } \mbox{ and } \mbox{ } \mbox{ } a_k=\dfrac{(2k)!}{4^k (k!)^2(2k+1)}
\end{equation}
(see equation (25), \cite{MW14}), and
\begin{equation}\label{Cq}
C_q:=  \int_{0}^{L} J_0(\psi)^q \psi \,d\psi, \mbox{  } \mbox{ for } q=3 \mbox{ and }q\geq 5,
\end{equation} with
\begin{equation}\label{J0}
 J_0(x)=\sum_{k=0}^{\infty} \dfrac{(-1)^kx^{2k}}{2^{2k}(k!)^2}
\end{equation} being the $J_0$ Bessel function.
In the Appendix, we perform a numerical investigation on the value of the constant $C$; more precisely, to obtain a precision of $1.0\times10^{-4}$, it is sufficient to sum the terms in (\ref{C e a}) until $q=20$, obtaining the value
\begin{equation}\label{numerino}
C=32\pi \sum_{k=1}^{20} a_k C_{2k+1}= 32\pi \times 0.1182.
\end{equation}
The constants $C_q$ are obtained by numerical integration, whereas for $C_3$ the exact value is  computed in \cite{MW11} and it is given by $$C_3=\dfrac{2}{\pi \sqrt{3}}=0.3676.$$

\begin{remark}
	It is easily seen that $50\%$ of the contribution of the sum in (\ref{numerino}) comes from the first term, which is $0.0613$. Moreover, the sum of the first and  second term is $0.0860$ (see Appendix), and thus, $80\%$ of the variance for the defect is explained by the third and fifth chaoses alone.
\end{remark}

Summing up, for $u=0$ the leading term in equation (\ref{var nodale}) disappears and we have to use the higher order approximation to find that
\begin{equation}
     \text{Var}\left( \mathcal{L}_{2}(A_{u=0}(f_{\ell };\mathbb{S}^{2})) \right) =  (4\pi)^2 \frac{0.0188}{\ell^2} + o(\frac{1}{\ell^2}) \text{ .}
\end{equation}

\begin{figure}[H]
  \begin{center}
    \includegraphics[width=1\textwidth,angle=0]{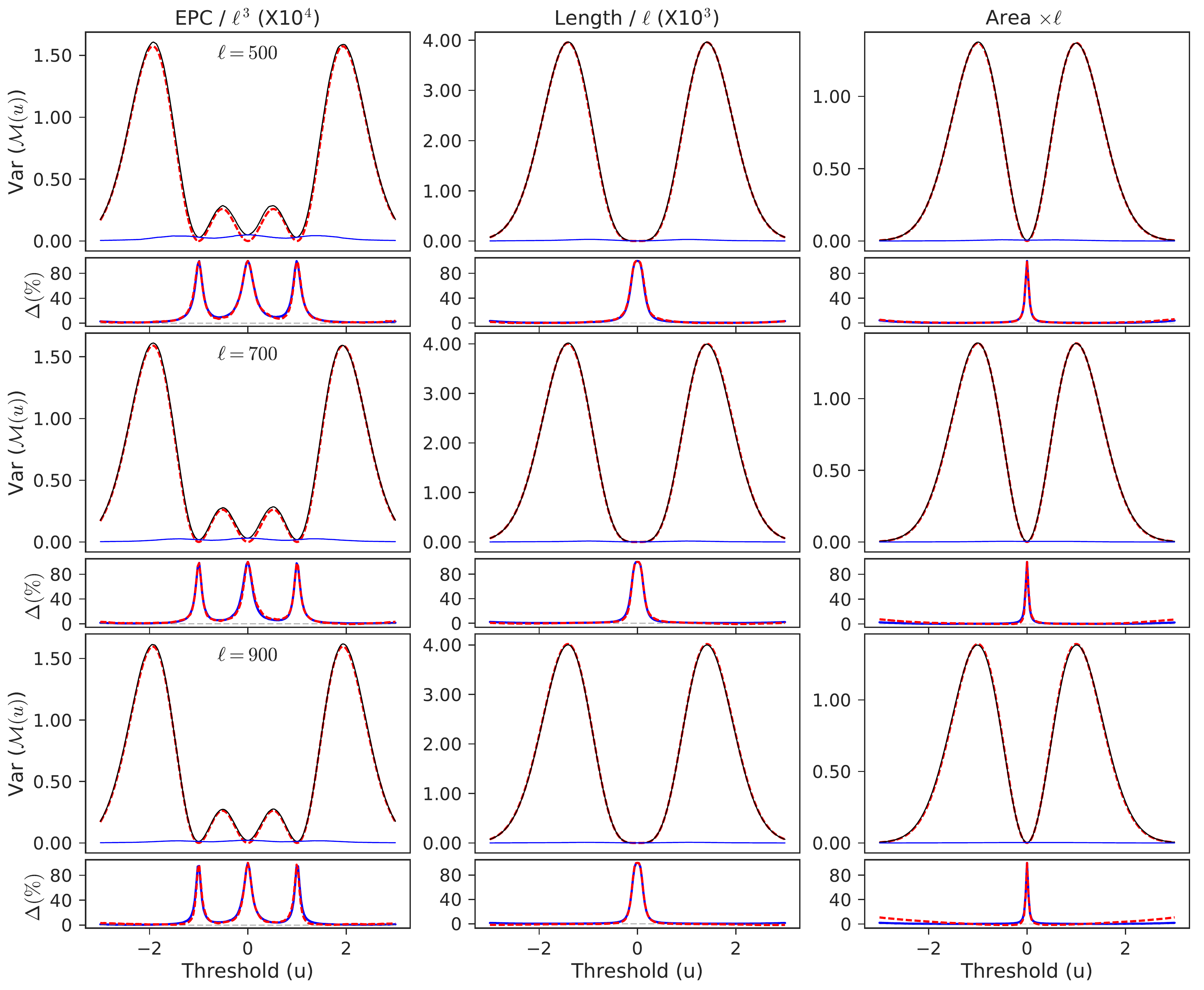}
\caption{Analytical vs simulation variance comparison. The red
  dash-dash curves are for analytical predictions; the black curves
  are the variances from 1000 simulations centered at the analytical
  predictions of the expectation values $proj[\mathcal{L_k}][0]$. The
  blue curves are variances of the simulations after subtracting the
  second order chaos term, i.e., 
  $proj[\mathcal{L_k}][0]+proj[\mathcal{L_k}][2]$ from each simulation. From the left to right, the theoretical predictions are: $\frac{1}{128 \pi^3} u^2(u^2-1)^2 e^{-u^{2}}$, $\frac{1}{128}u^4e^{-u^2}$, $\frac{1}{8\pi}u^2e^{-u^2}$.
  In the difference panels, the dash-dash red curves show the percentage difference between model and
  simulation variances, while the overlapping solid blue curves are the percentage
  ratio of the blue curve to the black in the main panel. The upper panel is for $\ell=500$, the middle for  for $\ell=700$ and the lower for $\ell=900$. \label{fig:var2ell}}
\end{center}
\end{figure}

In all the above equations, normalizing the area by $4\pi$ divides out the $16\pi^2$ term; this is the normalization that we shall adopt in the tables to follow in Section \ref{sec:numerical} (Tables \ref{tab:genus_mean}, \ref{tab:length_mean_rms}, \ref{tab:area_mean_rms} and Figures \ref{fig:m1ell}, \ref{fig:var2ell}).

\vspace{0.5cm}

\emph{\bf b) (Half) The Boundary Length ($k=1$)} %
Let us now consider the boundary length of excursion regions. To compute the expected value, it is enough to exploit the Gaussian Kinematic Formula; as before, note that we shall normalize by $4\pi$ in the simulations (see Table \ref{tab:length_mean_rms}) so that we obtain 
\begin{eqnarray*}
\mathbb{E}[\mathcal{L}_{1}(A_{u}(f_{\ell }; \mathbb{S}^2))]&=&   \mathtt{Proj}[\mathcal{L}_{1}(A_{u}(f_{\ell };\mathbb{S}^{2}))|0] =\left\{
\frac{\lambda _{\ell }}{2}\right\} ^{1/2}\sqrt{\frac{\pi }{8}}\left[
H_{0}(u)\phi (u)\right] \int_{\mathbb{S}^2} H_0(f_\ell(x))\,dx\\
&=& \frac{\sqrt{\ell(\ell+1)}}{4\sqrt{2}}e^{-u^2/2}4\pi \text{ .}
\end{eqnarray*}
Likewise, using results in \cite{Rossi}, \cite{W09}, \cite{MRW17}, we have for the leading stochastic term
\begin{equation*}
\mathtt{Proj}[\mathcal{L}_{1}(A_{u}(f_{\ell };\mathbb{S}^{2}))|2]=\frac{1}{2}%
\left\{ \frac{\lambda _{\ell }}{2}\right\} ^{1/2}\sqrt{\frac{\pi }{8}}\left[
H_{1}^{2}(u)\phi (u)\right] \int_{\mathbb{S}^{2}}H_{2}(f_{\ell }(x))dx\text{
;}
\end{equation*}
and using again (\ref{Varh2}) the variance can easily be seen to be  
\begin{eqnarray}\label{nodal lenght}
    \text{Var} \left( \mathcal{L}_{1}(A_{u}(f_{\ell };\mathbb{S}^{2})) \right) &=& 
    \frac{1}{4}\left\{ \frac{\lambda _{\ell }}{2}\right\}\frac{\pi }{8}\frac{u^4e^{-u^2}}{2\pi}\frac{16\pi^2}{\ell}+o(\frac{1}{\ell}) \\
    &=& \frac{\pi^2}{8}u^4e^{-u^2}(\ell+1)+o(\frac{1}{\ell}) \text{ .}
\end{eqnarray}

Again, in the simulations below (see Table \ref{tab:length_mean_rms}), normalizing the boundary length by $4\pi$ divides by a factor $16\pi^2$, leading (up to negligible terms) to a variance of order $\frac{\ell}{128}u^4e^{-u^2}$. \\\\
For $u=0$ the leading term in the previous expression disappears (the so-called Berry's cancellation phenomenon, see \cite{berry}, \cite{W09}) and the variance is of smaller order; more precisely, we have that (\cite{W09})
\begin{equation}\label{log}
 \text{Var} \left( \mathcal{L}_{1}(A_{u}(f_{\ell };\mathbb{S}^{2})) \right) =\frac{\log \ell}{128}+O(1) \text{ ;}
\end{equation}
(the same happens for shrinking subdomains of the sphere, see \cite{todino2}).
It is important to notice that the difference between the leading and remainder terms is here only of logarithmic order, and we hence expect a less precise approximation (in relative terms) in the simulations. On the other hand, it should also be noted that the variances at stake are much smaller than for $u \neq 0$, and then the absolute error in the simulations will turn out to be particularly small.\\
Hence, when $u=0$, the leading term in ($\ref{nodal lenght}$) disappears and the nodal length is asymptotic to the sample trispectrum, namely $$\mathcal{M}_\ell=-\dfrac{1}{4} \sqrt{\dfrac{\ell(\ell+1)}{2}} \dfrac{1}{4!} h_{\ell,4},$$
where $h_{\ell,4}=\int_{\mathbb{S}^2}H_4(T_\ell(x))\,dx$, which is logarithmic and hence we derive (\ref{log}). To be clear, as given in \cite{MRW17},
$$ \text{Var}(h_{\ell;4})=4! (4\pi)^2 \int_{0}^{1} P_\ell^4(t) \,dt \sim 576 \frac{\log \ell}{\ell^2}+O(\frac{1}{\ell^2})$$ and, setting $L:=\ell+\frac{1}{2}$, since
$$\lim_{\ell \rightarrow \infty} \ell^2 \int_{0}^{1} P_\ell^4(t)\, dt = \int_{0}^{L} J_0^4(\psi) \psi \,d\psi +O\bigg(\dfrac{1}{\ell}\bigg),$$ 
 we compute the last integral numerically, exploiting Matlab. We report some values in the table below.
\begin{center}
\begin{tabular}{|c|c|}
	\hline
	$L$ &\textbf{	$\int_{0}^{L} J_0^4(\psi)\psi\,d\psi$ } \\
	\hline 500 & 1.2420\\
	\hline 600&  1.2696\\
	\hline  1000 &  1.3475 \\
	\hline 1200 & 1.3751\\
	\hline 1500 & 1.4091 \\
	\hline 2000 & 1.4528 \\
	\hline
\end{tabular}
\end{center}

More explicitly, it was shown in \cite{MW14} that 

\[
\int_{0}^{1} P_\ell^4(t) \,dt \sim \dfrac{3}{2\pi^2} \dfrac{\log \ell}{\ell^2};
\]
to find a better approximation, we evaluate numerically the constant
\begin{equation}\label{1}
    \lim_{\ell \rightarrow \infty} \big[\int_{0}^{L} J_0^4(\psi) \psi \, d\psi -\dfrac{3}{2\pi^2} \log \ell \big]=0.297,
\end{equation}
see the Appendix for some analytic results. Hence, we conclude that, up to smaller order terms
\begin{equation}
\begin{split}
    \text{Var}(h_{\ell;4})&\sim 4! (4\pi)^2 \dfrac{1}{\ell^2} \bigg\{ \frac{3}{2\pi^2} \log \ell +0.297 \bigg\}= 4!16\pi^2 \dfrac{1}{\ell^2} \dfrac{3}{2\pi^2} \bigg\{ \log \ell+0.297\dfrac{2\pi^2}{3} \bigg\}\\&=576 \dfrac{1}{\ell^2} \bigg\{ \log \ell+1.9542 \bigg\}.
\end{split}
\end{equation}

Then, the variance of the scaled sample trispectrum $\mathcal{M}_{\ell}$ is asymptotically given by
$$\text{Var}(\mathcal{M}_{\ell}) \sim \dfrac{1}{16} \dfrac{\ell(\ell+1)}{2} \dfrac{1}{4!^2} 576 \dfrac{1}{\ell^2} \bigg\{ \log \ell+1.9542 \bigg\}= \dfrac{1}{32} \bigg\{ \log \ell+1.9542 \bigg\}$$

Finally, let us recall that these results, as in \cite{W09}, \cite{Rossi}, \cite{MRW17} and \cite{todino2} refer to the boundary length, not to the first Lipschitz-Killing curvature; there is hence a difference of a factor 2 in the expected value, and a factor 4 in the variance. The values in the Table \ref{tab:length_mean_rms} refer to the Lipschitz-Killing curvature, hence they have been normalized accordingly.

\vspace{0.5cm}

\emph{\bf c) Euler-Poincar\'e characteristic ($k=0$)} %
The Euler-Poincar\'e characteristic (EPC) for random spherical harmonics was investigated by \cite{CMW16b}, \cite{CM18} among others, where the following expressions are given for the expected value and the second chaotic component:

\begin{equation}
\mathtt{Proj}[\mathcal{L}_{0}(A_{u}(f_{\ell };\mathbb{S}^{2}))|0]=\left\{ 
\frac{\lambda _{\ell }}{2}\right\} \left[ H_{1}(u)\phi (u)\right] \frac{1}{%
2\pi }\int_{\mathbb{S}^{2}}H_{0}(f_{\ell }(x))dx+2\left\{ 1-\Phi (u)\right\} 
\text{ ,}
\end{equation}%
\begin{equation}\label{2comp}
\mathtt{Proj}[\mathcal{L}_{0}(A_{u}(f_{\ell };\mathbb{S}^{2}))|2]=\frac{1}{2}%
\left\{ \frac{\lambda _{\ell }}{2}\right\} \left[ H_{2}(u)H_{1}(u)\phi (u)%
\right] \frac{1}{2\pi }\int_{\mathbb{S}^{2}}H_{2}(f_{\ell }(x))dx+O_{p}(1)%
\text{ .}
\end{equation}%

All the EPC equations are normalized by $4\pi$ in the simulations, hence the $16\pi^2$ term is divided out.

Given these results, \cite{CM18} showed that the variances of LKCs
are dominated by the variance of the second order Wiener chaos; indeed, for
the Euler-Poincar\'e characteristic the expected value and variance are given, for an interval $I \subset \mathbb{R}$, by
$$\mathbb{E}[\chi(A_I(f_\ell;\mathbb{S}^2))]= \frac{2}{\sqrt{2\pi}} \ell(\ell+1) \int_I (t^2-1) e^{-t^2/2} \,dt $$
\begin{align*}
\text{Var}[\chi (A_{I}(f_{\ell }; {\cal \mathbb{S}}^2) )]
&=\frac{\ell^3}{8 \pi } 
\Big[   \int_{I}  (-t^4+4 t^2 -1) e^{-\frac {t^2} 2 } d t \Big]^2+O(\ell^{5/2}),  
\end{align*}

\noindent and in particular for semi-intervals of the form $I=[u,\infty)$ one obtains
$$
\mathbb{E}[\chi(A_u(f_\ell;\mathbb{S}^2))]= \sqrt{\frac{2}{\pi}} e^{-u^2/2} u \dfrac{\ell(\ell+1)}{2} +2 [1-\Phi(u)]$$
\begin{align*}
\text{Var}[\chi (A_{u}(f_{\ell }; {\cal \mathbb{S}}^2) )]&= \frac{\ell^3}{8 \pi }  e^{- u^2  }   (u-u^3)^2  +O(\ell^{5/2})\\
&=\frac{\ell^3 }{4 } 
 \frac{e^{-u^2}}{2\pi} [H_3(u)+2H_1(u) ]^2  +O(\ell^{5/2}).
\end{align*}

Note that, after normalizing the Lipschitz-Killing curvatures by their expected value, their relative variances converge to zero as the frequency increases, so that relative fluctuations become negligible on
small scales (Tables \ref{tab:genus_mean}, \ref{tab:length_mean_rms}, \ref{tab:area_mean_rms}).

\section{Characterization of Critical Points for Random Spherical Harmonics}\label{third sec}

As a further tool of investigation, we shall consider in this paper also the behaviour of critical points for random spherical harmonics, which has recently been fully characterized by \cite{CMW16, CW17, CM18b}, among others.

More precisely, by definition critical points, extrema and saddles are, respectively, given by:

\begin{equation*}
\mathcal{N}^{c}(f_\ell;u )=\mathcal{N}_{u}^{c}(f_{\ell })=\#\{x\in {\cal \mathbb{S}}^2
:f_{\ell }(x)\geq u,\nabla f_{\ell }(x)=0\},
\end{equation*}
\begin{equation*}
\mathcal{N}^{e}(f_\ell;u)=\mathcal{N}_{u}^{e}(f_{\ell })=\#\{x\in {\cal \mathbb{S}}^2
:f_{\ell }(x)\geq u,\nabla f_{\ell }(x)=0,\text{det}(\nabla ^{2}f_{\ell
}(x))>0\},
\end{equation*}
\begin{equation*}
\mathcal{N}^{s}(f_\ell;u)=\mathcal{N}_{u}^{s}(f_{\ell })=\#\{x\in {\cal \mathbb{S}}^2
:f_{\ell }(x)\geq u,\nabla f_{\ell }(x)=0,\text{det}(\nabla ^{2}f_{\ell
}(x))<0\}.
\end{equation*}%
 where we used $a=c,e,s$ to label critical points, extrema and saddles respectively. \\


We now recall the following results on the expectations and variances:

\medskip

For every interval $u \in \mathbb{R}$, we have, as $\ell \rightarrow \infty$,
\begin{equation*}
\mathbb{E}[\mathcal{N}_{u}^{a}(f_{\ell })] =\frac{2}{\sqrt{3}} \ell(\ell+1) \int_{u}^{\infty}\pi _{1}^{a}(t)\,dt+O(1),
\end{equation*}
where $a=c,e,s$ and for the density
functions

\begin{align}  \label{eqn:exp_crit}   
  \pi _{1}^{c}(t)&=\frac{\sqrt{3}}{\sqrt{8\pi
                   }}(2e^{-t^{2}}+t^{2}-1)e^{-\frac{                   
                   t^{2}}{2}}, \\
  \pi _{1}^{e}(t)&=\frac{\sqrt{3}}{\sqrt{2\pi
                   }}(e^{-t^{2}}+t^{2}-1)e^{-\frac{                   
                   t^{2}}{2}},\\
  \pi _{1}^{s}(t)&=\pi _{1}^{c}(t)-\pi
                   _{1}^{e}(t)=\frac{\sqrt{3}}{\sqrt{2\pi
                   }}e^{-\frac{3}{2}t^{2}}.
\end{align}

Similarly, for every $u \in \R$, as $\ell \rightarrow \infty $,
\begin{equation*}
{\text{Var}}(\mathcal{N}_{u}^{a}(f_{\ell }))=\ell^3 \left[
\int_{u}^{\infty}p_{3}^{a}(t)\,dt\right] ^{2}+O(\ell^{2}\log \ell),
\end{equation*}
where,

\begin{align*}
  p_{3}^{c}(t)&=\frac{1}{ \sqrt{8 \pi }}e^{-\frac{3}{2} t^{2}
                }[2-6t^{2}-e^{t^{2}}(1-4t^{2}+t^{4})], \\
  p_{3}^{e}(t)&=\frac{1}{ \sqrt{8 \pi }}e^{-\frac{3}{2} t^{2}
                }[1-3t^{2}-e^{t^{2}}(1-4t^{2}+t^{4})],\\
  p_{3}^{s}(t)&=\frac{1}{ \sqrt{8 \pi }}(1-3t^{2})e^{-\frac{3}{2}%
                t^{2}}. \\
  \label{eqn:var_crit}
\end{align*}

The leading constants for the variances can be written more explicitly as
\begin{align}
  \left[ \int_{u}^{\infty }p_3^c(t)\,dt \right]^{2}&=\frac{1}{8 \pi} e^{- 3 u^2} u^2 (2+ e^{u^2}
                                                   (u^2-1))^2, \\
  \left[ \int_{u}^{\infty }p_3^e(t)\,dt \right]^{2}&= \frac{1}{8 \pi}  e^{- 3
                                                   u^2} u^2 (1+ e^{u^2} (u^2-1))^2,\\
  \left[ \int_{u}^{\infty }p_3^s(t)\,dt \right]^{2}&=\frac{1}{8 \pi}  e^{- 3 u^2} u^2.
\end{align}
Note that also in this case, the second component is the leading term of the expansion and it is important to stress how the leading terms in the variances cancel in all cases at the threshold $u=-\infty$; in other words, the variance is smaller when we focus on the total number of critical points (see \cite{CW17}). This is again a form of the so-called ``Berry's cancellation phenomenon", which we have also discussed earlier for the Lipschitz-Killing curvatures. Indeed, the behaviour of critical points and saddles can be shown to be dominated by the second order chaotic component, which takes the form (see \cite{CM18})
$$\frac{\lambda_\ell}{2} \bigg[\int_{u}^{\infty}p_3^a(t)\,dt\bigg] \frac{1}{2\pi} \int_{\mathbb{S}^2} H_2(f_\ell(x)) \,dx$$
and similarly for saddles. 

Because this second-order chaos component (and hence the leading term in the variance) vanishes at $u=-\infty,0$, the next component becomes of interest; it can be shown that this term is proportional to the fourth-order chaos, and indeed, for the total number of critical points, it holds that (see \cite{CW17})
\begin{equation*}
\mathbb{E}[\mathcal{N}_{-\infty}^{c}(f_{\ell })]=\frac{2\ell(\ell+1)}{\sqrt{3}}+O(1) \text{ , } {\text{Var}}(\mathcal{N}_{-\infty}^{c}(f_{\ell }))=\frac{\ell^2 \log \ell}{27 \pi^2} + O(\ell^2)\text{ ;}    
\end{equation*}
moreover, it is also possible to consider separately extrema (minima and maxima) and saddles, yielding
\begin{equation*}
\mathbb{E}[\mathcal{N}_{-\infty}^{e}(f_{\ell })]=\frac{\ell(\ell+1)}{\sqrt{3}}+O(1) \text{ , }{\text{Var}}(\mathcal{N}_{-\infty}^{e}(f_{\ell }))=\frac{\ell^2 \log \ell}{4 \times 27 \pi^2} + O(\ell^2)\text{ ,}   
\end{equation*}
and
\begin{equation*}
\mathbb{E}[\mathcal{N}_{-\infty}^{s}(f_{\ell })]=\frac{\ell(\ell+1)}{\sqrt{3}}+ O(1) \text{ , }{\text{Var}}(\mathcal{N}_{-\infty}^{s}(f_{\ell }))=\frac{\ell^2 \log \ell}{4 \times 27 \pi^2}+ O(\ell^2) \text{ .}   
\end{equation*}

\section{On Correlations}\label{corr sec}

The results presented in the previous sections can be summarized as follows:

1) For general threshold $u \neq 0$, the fluctuations around the proper expected values for the area, the boundary length and the Euler-Poincar\'e characteristic of excursion regions is dominated by a single stochastic term, which is proportional to the so-called second order Wiener chaos; namely $h_{\ell;2}=\int_{\mathbb{S}^2} H_2(f_{\ell}(x))\,dx$.

2) At $u=0$, this term is disappearing; the boundary length is then dominated by the fourth-order chaos, i.e., a single term which is proportional to $h_{\ell;4}=\int_{S^2} H_4(f_{\ell}(x))dx$. For the excursion area and the Euler-Poincar\'e, this term is disappearing as well and lower order terms are dominant.

3) Likewise, the critical points above general threshold levels $u$ are dominated by a single term, proportional to $h_{\ell;2}$; this term disappears for $u=-\infty$, where the total number of critical points is dominated by a single term proportional to $h_{\ell;4}$.  

Note also that the variance of $h_{\ell;2}$ is of order $O(\frac{1}{\ell})$, the variance of $h_{\ell;4}$ is of order $O(\frac{\log\ell}{\ell^2})$, and the variance of all other chaoses (for $q=3,5,6,7...$) is of order $O(\frac{1}{\ell^2}$). As a consequence, we expect almost perfect correlation for all statistics which are dominated by $h_{\ell;2}$; some correlation (but not too strong, given the logarithmic rate) for statistics dominated by $h_{\ell;4}$; no correlation for statistics which are dominated by chaoses of different order. These conjectures are indeed very well confirmed by the numerical evidence that we shall present in the Section below (Fig. \ref{fig:corr}).

\section{Numerical results}\label{sec:numerical}

In this section we describe the comparison of the analytical
results outlined in the previous sections to the corresponding
results from simulations.  In order to implement this comparison, we generated 1000 Gaussian
realizations of random spherical eigenfunctions/spherical harmonics for different values of the multipoles $\ell$, ranging from $\ell=100$ to $\ell=900$. These values for the multipoles $\ell$ are representative of the resolution which can be currently achieved by satellite experiments such as Planck (see \cite{planck2018missionpaper}); for instance, these eigenfunctions could be taken to be the spherical Fourier component of a simulated CMB map, according to a standard routine provided by the \healpix\cite{healpix} package. The simulations algorithms are described more fully in the subsection to follow.

\subsection*{Simulations and Algorithm}

We used the \healpix \emph{synfast} routine to simulate a Gaussian realization map starting from a given power-spectrum. In practice, we used the so-called best-fit Planck power spectrum to generate the maps, and then we extracted the multipoles to focus on, normalizing their variance to unity. Of course, our results are independent from the choice of the input power spectrum, and indeed it would be possible to generate directly the single eigenfunctions at a given multipole.

A single multipole map $f_\ell(x)$ is obtained by using the \healpix \emph{alm2map} routine, after having extracted the proper subset of coefficients $a_{\ell m}$. In all cases the map resolution parameter $N_{side}$ is set to twice the value of the corresponding multipole. As mentioned earlier each map is normalized to have unit variance.


It is very important to notice that each functional is normalized ``per unit area", i.e., all the reported values have been standardized dividing by $4\pi$. Both the expected values and the variances are affected in the obvious way.

We compute  the three Minkowski functionals, which are equivalent to the LKCs up to constant factors, and critical point counts from these normalized multipole maps.  
In short, the area, i.e. the first Minkowski functional, is simply computed by evaluating
the number of pixels above a certain threshold.  The perimeter length, the second Minkowski functional, is computed by tracing isocontour lines in pixel space. For a sufficiently high-resolution map, pixels around isocontour
lines have different signs relative to the contour line, after normalizing the lines to zero. To measure the length of these lines, sets of four pixels are compared; when at least two of them have different signs, the locations where the contour line enters and exits these sets of pixels are determined and the length is iteratively calculated by standard dot product. 
For the Euler-Poincar\'{e} characteristic, the third Minkowski functional, we used the Fortran implementation of the algorithms described in Appendix G of
\cite{Gay2012_NGPeaks} (see also \cite{FMHM15}). This algorithm is based on the Gauss-Bonnet
theorem - where the Euler characteristic of a region is obtained by
integrating the curvature over the boundary surface. Given we are working on a pixelized surface, the surface curvature of
an excursion region can be thought of as
concentrated in the corners of the pixels that
are at the boundary between the pixels above and below the threshold. This
is true as any continuous deformation of the region conserve the
topology. What is needed
is, therefore, to devise a strategy that assigns appropriate curvature weights
for each boundary grid vertex - Appendix G of
\cite{Gay2012_NGPeaks} explains in more detail the strategy used in the Fortran code. Once the weights are assigned, the sum of the weights over all the vertices gives us the Euler characteristic of the excursion set.


Our detailed investigation using different algorithms to compute the
Euler-Poincar\'{e} characteristic showed that for a map defined at a
given $N_{\rm side}$, the maximum multipole for which a percent
numerical accuracy can be obtained is $\ell_{max} \sim N_{\rm
  side}/3$. While it would be
possible to cover larger values, we do not believe this is
essential for our purpose in this paper.

\subsection*{Results}


We first proceed to report in Tables \ref{tab:genus_mean}, \ref{tab:length_mean_rms}, \ref{tab:area_mean_rms}, a numerical comparison of the expected values computed on simulated maps and their analytical predictions, and likewise, Monte Carlo estimates of root mean squared errors and their analytical predictions. We stress that the fit is truly remarkable: the percentage errors are smaller than 1\% for most non-zero values of the threshold parameter. It should be recalled here that the analytical predictions for expected values are exact, while for the variances we are only giving the leading term in a series of positive addends; for non-zero values of $u$, the neglected terms in the variance (as mentioned earlier) are a factor $\ell$ smaller than the leading one, as mirrored in the simulations. 

\begin{table}[H]
\centering
\begin{tabular}{|c||c|c|c|c|c|c|c|c|c|c|}
\multicolumn{11}{c}{} \\
\multicolumn{11}{c}{\textbf{Expected Values}} \\ 
\hline 
& \multicolumn{2}{c|}{$\ell=100$} & \multicolumn{2}{c|}{$\ell=300$} & \multicolumn{2}{c|}{$\ell=500$} & \multicolumn{2}{c|}{$\ell=700$} & \multicolumn{2}{c|}{$\ell=900$}\\ 
 \hline
Threshold & Sim & Model & Sim & Model & Sim & Model & Sim & Model & Sim & Model\\ 
 \hline
\textbf{-3.0} & -10.28 & -10.53 & -95.60 & -95.38 & -267.70 & -264.88 & -525.56 & -519.01 & -868.04 & -857.79\\
\textbf{-1.5} & -155.50 & -156.00 & -1397.28 & -1395.90 & -3886.18 & -3872.59 & -7628.09 & -7586.09 & -12629.25 & -12536.39\\
\textbf{0.0} & 0.13 & 0.08 & 1.07 & 0.08 & 0.25 & 0.08 & -1.13 & 0.08 & 1.16 & 0.08\\
\textbf{1.5} & 155.62 & 156.16 & 1397.72 & 1396.05 & 3886.10 & 3872.75 & 7627.40 & 7586.25 & 12629.65 & 12536.55\\
\textbf{3.0} & 10.44 & 10.69 & 95.96 & 95.54 & 268.15 & 265.04 & 525.77 & 519.17 & 868.19 & 857.95\\
\hline 
\multicolumn{11}{c}{} \\
\multicolumn{11}{c}{\textbf{Standard Deviation}} \\ 
\hline 
& \multicolumn{2}{c|}{$\ell=100$} & \multicolumn{2}{c|}{$\ell=300$} & \multicolumn{2}{c|}{$\ell=500$} & \multicolumn{2}{c|}{$\ell=700$} & \multicolumn{2}{c|}{$\ell=900$}\\ 
 \hline
Threshold & Sim & Model & Sim & Model & Sim & Model & Sim & Model & Sim & Model\\ 
 \hline
\textbf{-3.0} & 4.23 & 4.26 & 21.63 & 22.05 & 46.80 & 47.39 & 77.76 & 78.46 & 119.51 & 114.36\\
\textbf{-1.5} & 10.02 & 9.74 & 49.76 & 50.33 & 107.02 & 108.19 & 175.20 & 179.14 & 275.18 & 261.11\\
\textbf{0.0} & 4.04 & - & 11.54 & - & 19.76 & - & 26.76 & - & 34.43 & -\\
\textbf{1.5} & 10.08 & 9.74 & 49.80 & 50.33 & 106.39 & 108.19 & 175.47 & 179.14 & 275.08 & 261.11\\
\textbf{3.0} & 4.26 & 4.26 & 21.74 & 22.05 & 46.87 & 47.39 & 77.54 & 78.46 & 120.12 & 114.36\\
\hline 
\end{tabular} \\
\caption{Expected Values and Standard Deviation of the Euler-Poincar\'e characteristic. The theoretical expressions are given in Tables \ref{table:lkcu} and \ref{table:lkc0} in the Appendix, (the EPC corresponds to $\mathcal{L}_0(A_u(f_\ell; \mathbb{S}^2))$; the reported values are normalized dividing by $4\pi$. Theoretical expressions at $u=0$ have yet to be determined.}
\label{tab:genus_mean}
\end{table}

\begin{table}[H]
\centering
\begin{tabular}{|c||c|c|c|c|c|c|c|c|c|c|}

\multicolumn{11}{c}{} \\ 
\multicolumn{11}{c}{\textbf{Expected Values}} \\ 
\hline 
& \multicolumn{2}{c|}{$\ell=100$} & \multicolumn{2}{c|}{$\ell=300$} & \multicolumn{2}{c|}{$\ell=500$} & \multicolumn{2}{c|}{$\ell=700$} & \multicolumn{2}{c|}{$\ell=900$}\\ 
 \hline
Threshold & Sim & Model & Sim & Model & Sim & Model & Sim & Model & Sim & Model\\ 
 \hline
\textbf{-3.0} & 0.193 & 0.197 & 0.587 & 0.590 & 0.977 & 0.983 & 1.347 & 1.376 & 1.696 & 1.768\\
\textbf{-1.5} & 5.736 & 5.768 & 17.225 & 17.246 & 28.667 & 28.724 & 39.965 & 40.202 & 51.077 & 51.681\\
\textbf{0.0} & 17.764 & 17.766 & 53.101 & 53.121 & 88.417 & 88.477 & 123.727 & 123.832 & 159.052 & 159.187\\
\textbf{1.5} & 5.732 & 5.768 & 17.232 & 17.246 & 28.667 & 28.724 & 39.961 & 40.202 & 51.079 & 51.681\\
\textbf{3.0} & 0.193 & 0.197 & 0.589 & 0.590 & 0.977 & 0.983 & 1.347 & 1.376 & 1.695 & 1.768\\

\hline 
\multicolumn{11}{c}{} \\ 
\multicolumn{11}{c}{\textbf{Standard Deviation}} \\ 
\hline 
& \multicolumn{2}{c|}{$\ell=100$} & \multicolumn{2}{c|}{$\ell=300$} & \multicolumn{2}{c|}{$\ell=500$} & \multicolumn{2}{c|}{$\ell=700$} & \multicolumn{2}{c|}{$\ell=900$}\\ 
 \hline
Threshold & Sim & Model & Sim & Model & Sim & Model & Sim & Model & Sim & Model\\ 
 \hline
\textbf{-3.0} & 0.088 & 0.089 & 0.150 & 0.153 & 0.193 & 0.198 & 0.226 & 0.234 & 0.266 & 0.265\\
\textbf{-1.5} & 0.634 & 0.647 & 1.083 & 1.119 & 1.406 & 1.444 & 1.662 & 1.709 & 2.007 & 1.937\\
\textbf{0.0} & 0.028 & 0.018 & 0.029 & 0.019 & 0.029 & 0.020 & 0.031 & 0.021 & 0.030 & 0.021\\
\textbf{1.5} & 0.635 & 0.647 & 1.083 & 1.119 & 1.402 & 1.444 & 1.661 & 1.709 & 2.007 & 1.937\\
\textbf{3.0} & 0.089 & 0.089 & 0.150 & 0.153 & 0.193 & 0.198 & 0.226 & 0.234 & 0.267 & 0.265\\
\hline 
\end{tabular}
\caption{Expected Values and Standard Deviation of half of the boundary length functional, i.e. $\mathcal{L}_1(A_u(f_\ell; \mathbb{S}^2))$; the reported values are normalized dividing by $4\pi$. Theoretical results are given in Tables \ref{table:lkcu} and \ref{table:lkc0} in the Appendix.} 
\label{tab:length_mean_rms}
\end{table}
\begin{table}[H]
\centering
\begin{tabular}{|c||c|c|c|c|c|c|c|c|c|c|}
\multicolumn{11}{c}{} \\ 
\multicolumn{11}{c}{\textbf{Expected Values}} \\ 
\hline 
& \multicolumn{2}{c|}{$\ell=100$} & \multicolumn{2}{c|}{$\ell=300$} & \multicolumn{2}{c|}{$\ell=500$} & \multicolumn{2}{c|}{$\ell=700$} & \multicolumn{2}{c|}{$\ell=900$}\\ 
 \hline
Threshold & Sim & Model & Sim & Model & Sim & Model & Sim & Model & Sim & Model\\ 
 \hline
\textbf{-3.0} & 0.9987 & 0.9987 & 0.9986 & 0.9987 & 0.9986 & 0.9987 & 0.9986 & 0.9987 & 0.9985 & 0.9987\\
\textbf{-1.5} & 0.9336 & 0.9332 & 0.9330 & 0.9332 & 0.9325 & 0.9332 & 0.9320 & 0.9332 & 0.9315 & 0.9332\\
\textbf{0.0} & 0.5000 & 0.5000 & 0.5000 & 0.5000 & 0.5000 & 0.5000 & 0.5000 & 0.5000 & 0.5000 & 0.5000\\
\textbf{1.5} & 0.0663 & 0.0668 & 0.0671 & 0.0668 & 0.0675 & 0.0668 & 0.0680 & 0.0668 & 0.0685 & 0.0668\\
\textbf{3.0} & 0.0013 & 0.0013 & 0.0014 & 0.0013 & 0.0014 & 0.0013 & 0.0014 & 0.0013 & 0.0015 & 0.0013\\

\hline 
\multicolumn{11}{c}{} \\ 
\multicolumn{11}{c}{\textbf{Standard Deviation}} \\ 
\hline 
& \multicolumn{2}{c|}{$\ell=100$} & \multicolumn{2}{c|}{$\ell=300$} & \multicolumn{2}{c|}{$\ell=500$} & \multicolumn{2}{c|}{$\ell=700$} & \multicolumn{2}{c|}{$\ell=900$}\\ 
 \hline
Threshold & Sim & Model & Sim & Model & Sim & Model & Sim & Model & Sim & Model\\ 
 \hline
\textbf{-3.0} & 0.0007 & 0.0007 & 0.0004 & 0.0004 & 0.0003 & 0.0003 & 0.0003 & 0.0003 & 0.0002 & 0.0002\\
\textbf{-1.5} & 0.0095 & 0.0097 & 0.0054 & 0.0056 & 0.0042 & 0.0043 & 0.0036 & 0.0037 & 0.0034 & 0.0032\\
\textbf{0.0} & 0.0014 & 0.0013 & 0.0005 & 0.0004 & 0.0003 & 0.0003 & 0.0002 & 0.0002 & 0.0002 & 0.0001\\
\textbf{1.5} & 0.0094 & 0.0097 & 0.0054 & 0.0056 & 0.0042 & 0.0043 & 0.0036 & 0.0037 & 0.0034 & 0.0032\\
\textbf{3.0} & 0.0007 & 0.0007 & 0.0004 & 0.0004 & 0.0003 & 0.0003 & 0.0003 & 0.0003 & 0.0002 & 0.0002\\
\hline 
\end{tabular}
\caption{Expected Values and Standard Deviation of the Area functional. The Area is normalized to unity, i.e., divided by $4\pi$. Theoretical expressions are summarized in Tables \ref{table:lkcu} and \ref{table:lkc0} in the Appendix. } 
\label{tab:area_mean_rms}
\end{table}
The analytic approximation for the variances in the case $u=0$ is slightly worse, in relative terms, but actually even better, in absolute terms. This was explained earlier in Section \ref{second sec}; in short, the variances at $u=0$ are an order of magnitude smaller than at other thresholds, because the leading term cancels, and new elements become dominant (the fourth-order chaos, in the case of the nodal length). Thus, focussing for instance on the boundary length, here the dominant term is larger than the neglected ones only by a logarithmic factor; as a consequence, variances tend to be underestimated (a similar phenomenon occurs for the total number of critical points, see below). In absolute terms, the discrepancy between simulations and analytic results for the nodal length is in the order of $10^{-3}/10^{-1}$, to be compared with expected values in the order of $10/10^{2}$, so that the relative error is in the order of $10^{-3}$. 

\begin{figure}
\begin{center}
  \includegraphics[width=1\textwidth,angle=0]{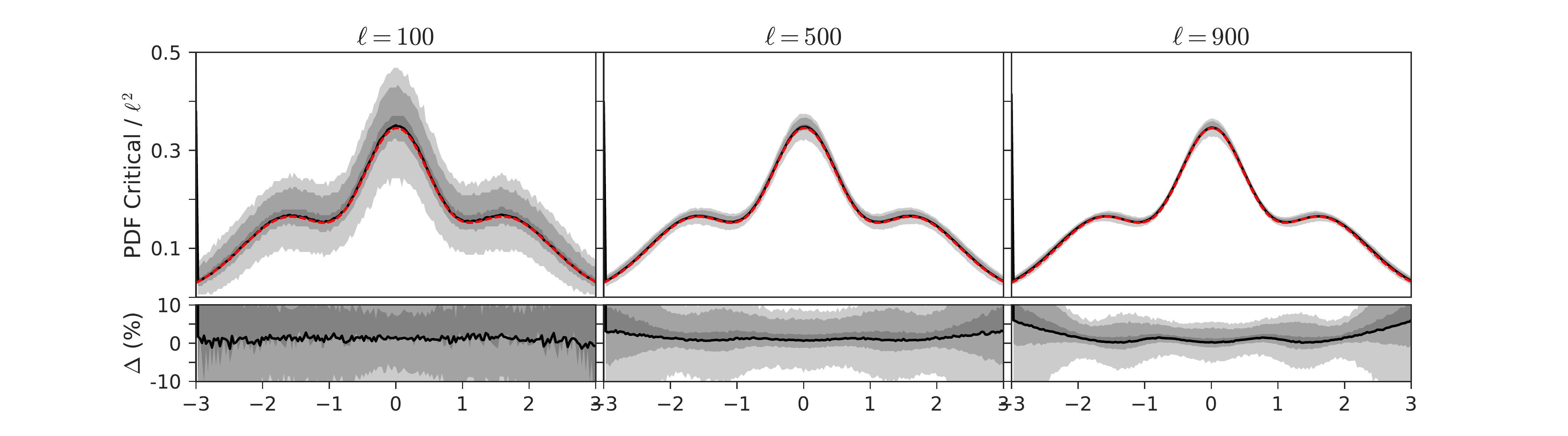}\\
  \includegraphics[width=1\textwidth,angle=0]{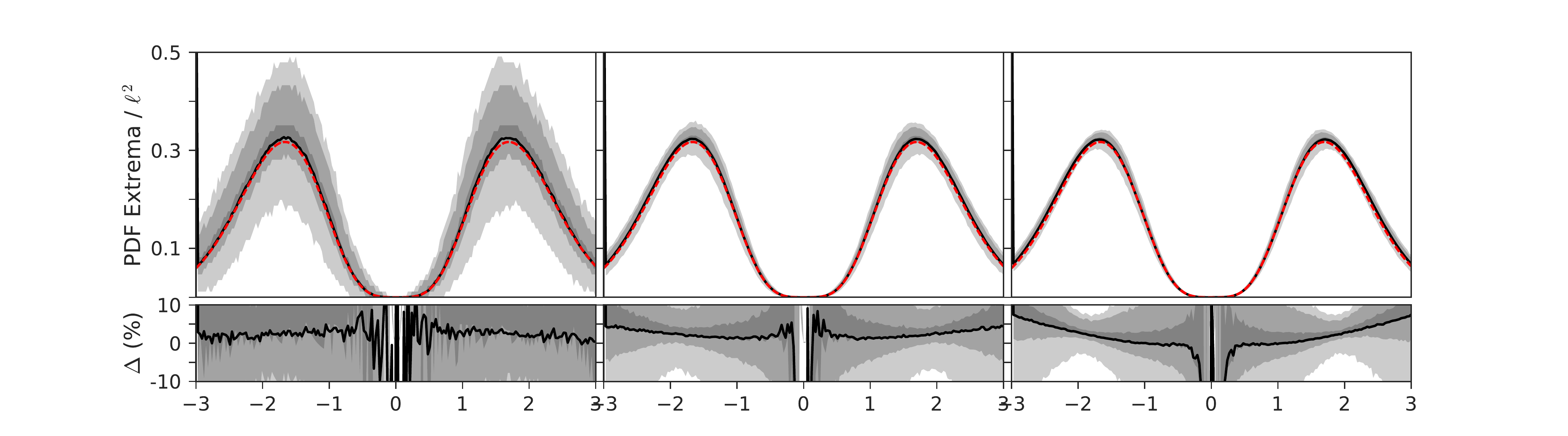}\\
  \includegraphics[width=1\textwidth,angle=0]{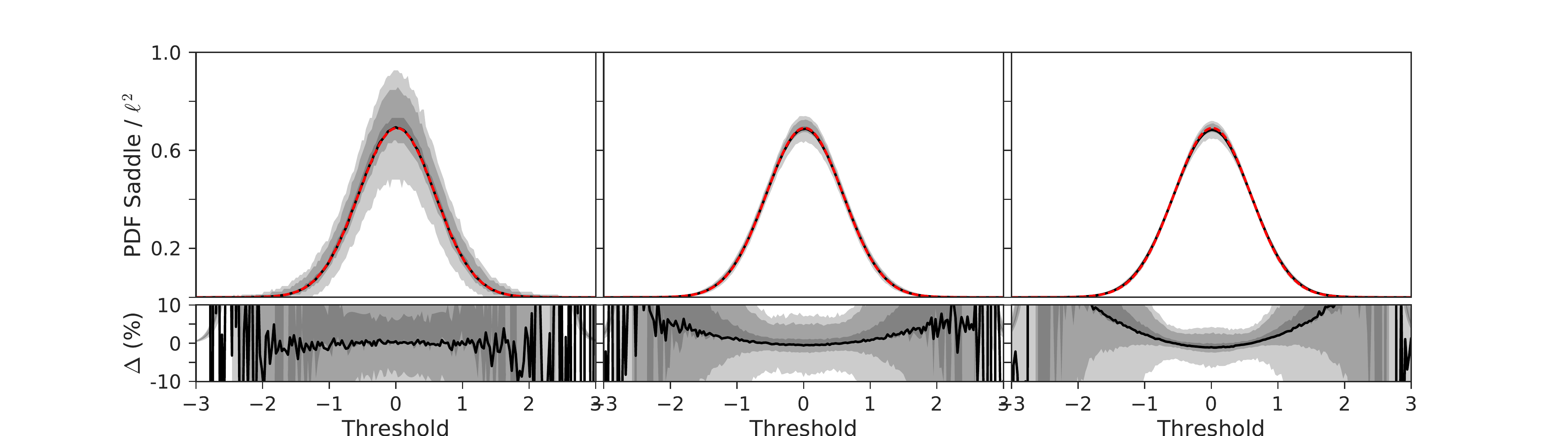}  
\caption{Comparison of expectation density of critical, extrema, and saddle points from theory and simulation. The red line shows the function $\frac{2}{\sqrt{3}} \pi_1^a(t)$, 
for $a=c,e,s$, from the top to the bottom, respectively.
The simulations curves are evaluated at multipoles $\ell=500,700,900$: Grey Shades are $68, 95$ and $99 \%$
  percentiles estimated from 1000 simulations. \label{fig:exp_crit}}
\end{center}
\end{figure}

\begin{figure}[H]
\begin{center}
  \includegraphics[width=1\textwidth,angle=0]{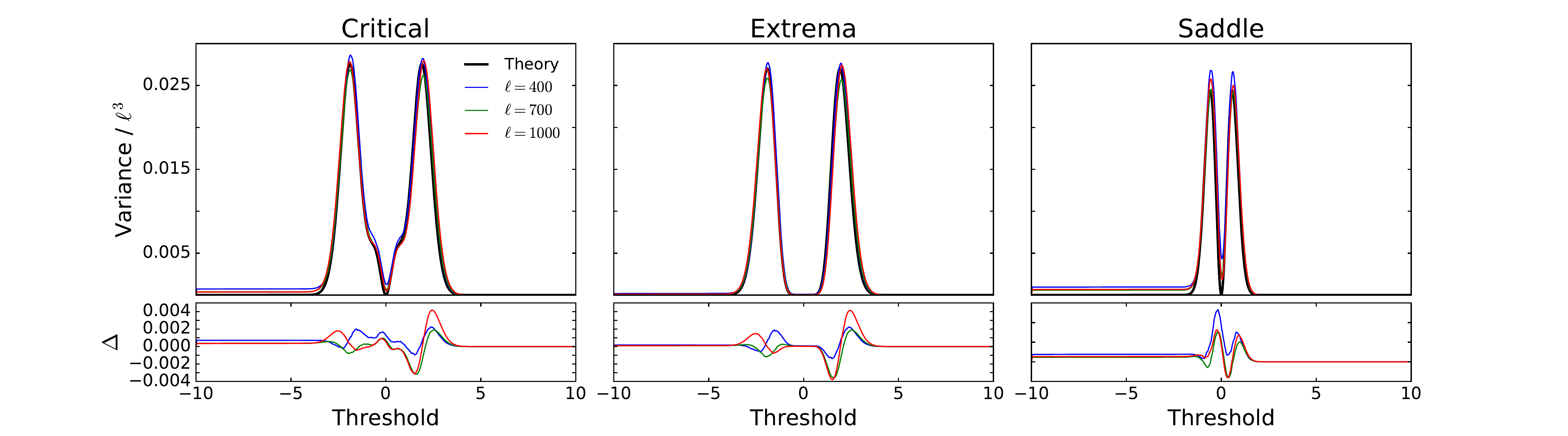}  
\caption{Comparison of Variance of critical, extrema, and saddle points from theory and simulation. The red line in the left panel represents the plot of the function $\frac{1}{8\pi} e^{-3u^2}u^2 (2+e^{u^2}(u^2-1))^2$, which yields the variance of critical values in the interval $(u,\infty)$. Likewise, in the other two panels it represents the variance of extrema and saddle points, i.e. the function $[\int_{u}^{\infty} p_3^a(t)\,dt]^2$ for $a=e$ in the middle panel and $a=s$ in the right one. 
  The legend shows the multipoles at
  which the simulation curves are evaluated: Grey Shades are $68, 95$ and $99 \%$
  percentiles estimated from 1000 simulations. \label{fig:var_crit}}
\end{center}
\end{figure}

The results for critical points (Tables \ref{tab:saddle_mean_rms}, \ref{tab:tot_critical_mean_rms} and Figures \ref{fig:exp_crit}, \ref{fig:var_crit}) are, in our view, equally impressive, with relative errors in the order of $10^{-3}/10^{-4}$, and absolute ones in the order of $10/10^2$, to be compared with expected values that run in the hundreds of thousands ($10^4,10^5$).

\begin{table}[H]
\centering
\begin{tabular}{|c||c|c|c|c|c|c|}
\hline 
& \multicolumn{2}{c|}{Critical} & \multicolumn{2}{c|}{Extrema} & \multicolumn{2}{c|}{Saddle}\\ 
 \hline
Threshold & Sim & Model & Sim & Model & Sim & Model\\ 
 \hline
\textbf{-3.0} & 0.0330 & 0.0318 & 0.0669 & 0.0635 & 0.0000 & 0.0000\\
\textbf{-1.5} & 0.1648 & 0.1641 & 0.3101 & 0.3061 & 0.0232 & 0.0221\\
\textbf{0.0} & 0.3471 & 0.3453 & 0.0000 & 0.0000 & 0.6855 & 0.6907\\
\textbf{1.5} & 0.1642 & 0.1635 & 0.3057 & 0.3017 & 0.0264 & 0.0253\\
\textbf{3.0} & 0.0332 & 0.0318 & 0.0672 & 0.0635 & 0.0000 & 0.0000\\

\hline 
\end{tabular}
\caption{Densities of the Expected Values of Critical Points, Extrema and Saddles. The simulation results are obtained for a bin width of $\Delta u=0.03$.   
The theoretical results are given by the values of $ \pi_1^a(t)$ defined in (\ref{eqn:exp_crit})-(34).
} 
\label{tab:saddle_mean_rms} 
\end{table}

\begin{table}[H]
\centering
\begin{tabular}{|c||c|c|c|c|c|c|}

\hline 

& \multicolumn{2}{c|}{Critical} & \multicolumn{2}{c|}{Extrema} & \multicolumn{2}{c|}{Saddle}\\ 
 \hline
$\ell$ & Mean & $\sigma$ & Mean & $\sigma$ & Mean & $\sigma$\\ 
 \hline
 
\textbf{100} & 11659.3300 & 52.7320 & 5830.6280 & 26.3011 & 5828.7020 & 26.4539\\
\textbf{300} & 104306.4320 & 165.2952 & 52151.1920 & 82.6017 & 52155.2400 & 83.0788\\
\textbf{500} & 289521.0780 & 275.0422 & 144729.9090 & 137.8649 & 144791.1690 & 139.0948\\
\textbf{700} & 567436.9110 & 371.9949 & 283565.7720 & 187.7662 & 283871.1390 & 189.0990\\
\textbf{900} & 937875.8670 & 479.1085 & 468449.3170 & 242.9301 & 469426.5500 & 247.7632\\

\hline 
\end{tabular} \\
\caption{Total number ($u=-\infty$) of critical points, extrema and saddles (Expected Values and their Standard Deviations). These mean values and standard deviations are computed from 1000 simulations. The theoretical expected value is $\frac{2}{\sqrt{3}} \ell(\ell+1)$ and the theoretical variance is $\frac{1}{3^3 \pi^2} \ell^2 \log \ell$, see Table \ref{table:cp} in the Appendix.}
\label{tab:tot_critical_mean_rms} 
\end{table}

\begin{figure}[H]

  \begin{center}
    \includegraphics[width=1\textwidth,angle=0]{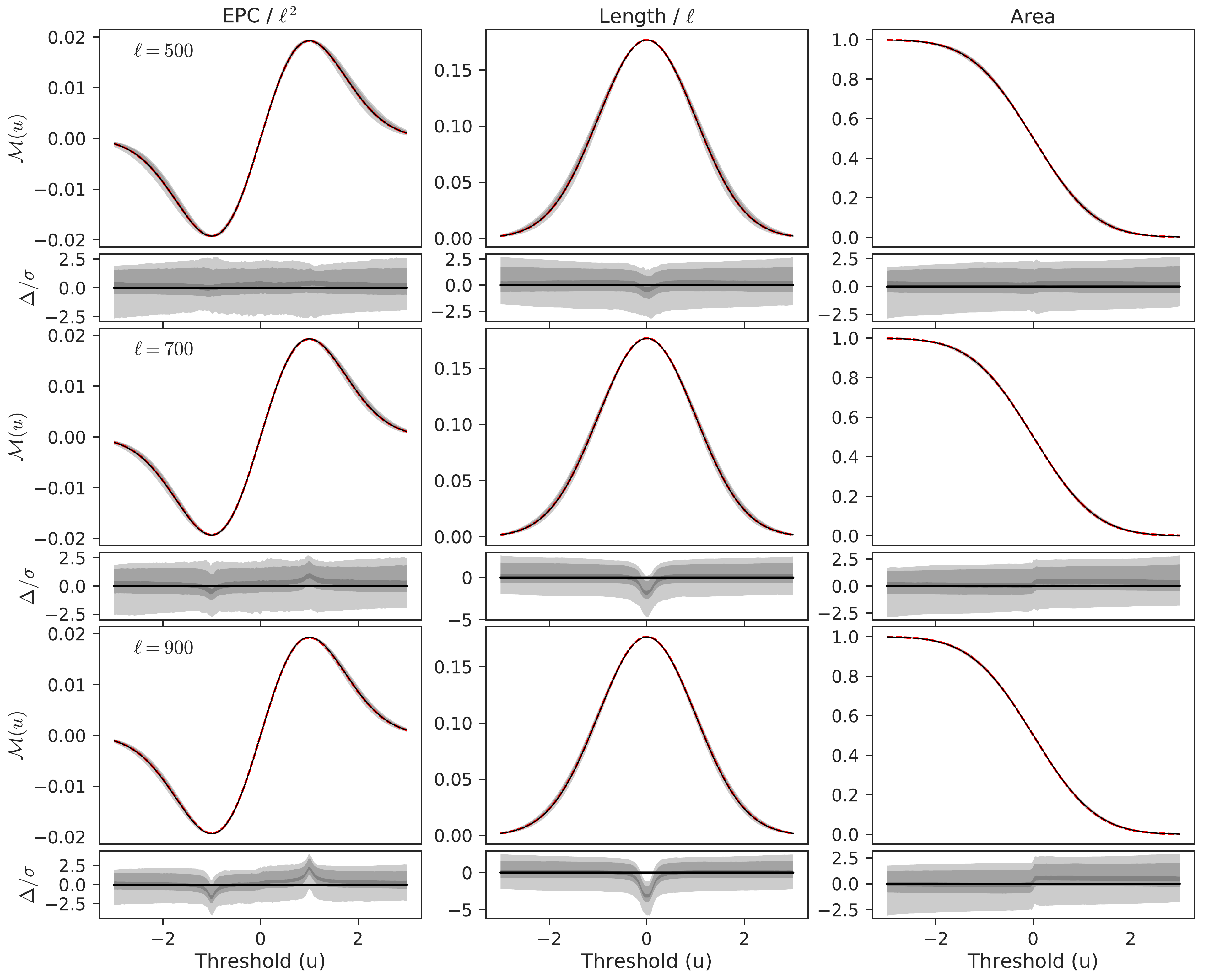}
\caption{Mean values and relative differences with theoretical predictions:
  grey and dash-dash red curves are for simulations and  
  analytical predictions, respectively. The rows from left to right are
  Euler-Poincar\'e characteristic, half of the boundary length and area,
  respectively. The theoretical predictions are, from left to the right: $\frac{1}{8\pi} \sqrt{\frac{2}{\pi}} e^{-u^2/2} u$, $\frac{1}{4\sqrt{2}}e^{-u^2/2}$, $1-\Phi(u)$, respectively.
  The upper panel is for $\ell=500$. the middle for $\ell=700$ and
  the lower panel for $\ell=900$. Grey Shades are $68, 95$ and $99 \%$
  percentiles estimated from 1000 simulations. \label{fig:m1ell}}
\end{center}
\end{figure}

To help visualization, we produced some plots that compare the analytic predictions with the realizations; more precisely, in Figure \ref{fig:m1ell} we compare the multipole space analytical
results (red curve) given in Section \ref{second sec} with that of the simulations (black curve - mean of the simulations). The $68\%,95\%$ and $99\%$ Confidence Intervals are shown from dark to light grey
bounds. From the top to the bottom rows, the figures show the plots of the results
corresponding to multipoles $\ell=500,700,900$. We stress that our
fit is extremely accurate even at low multipole values; we also note the improved concentration around the expected values at higher-multipoles.

\begin{figure}[H]
\begin{center}
  \includegraphics[width=1\textwidth,angle=0]{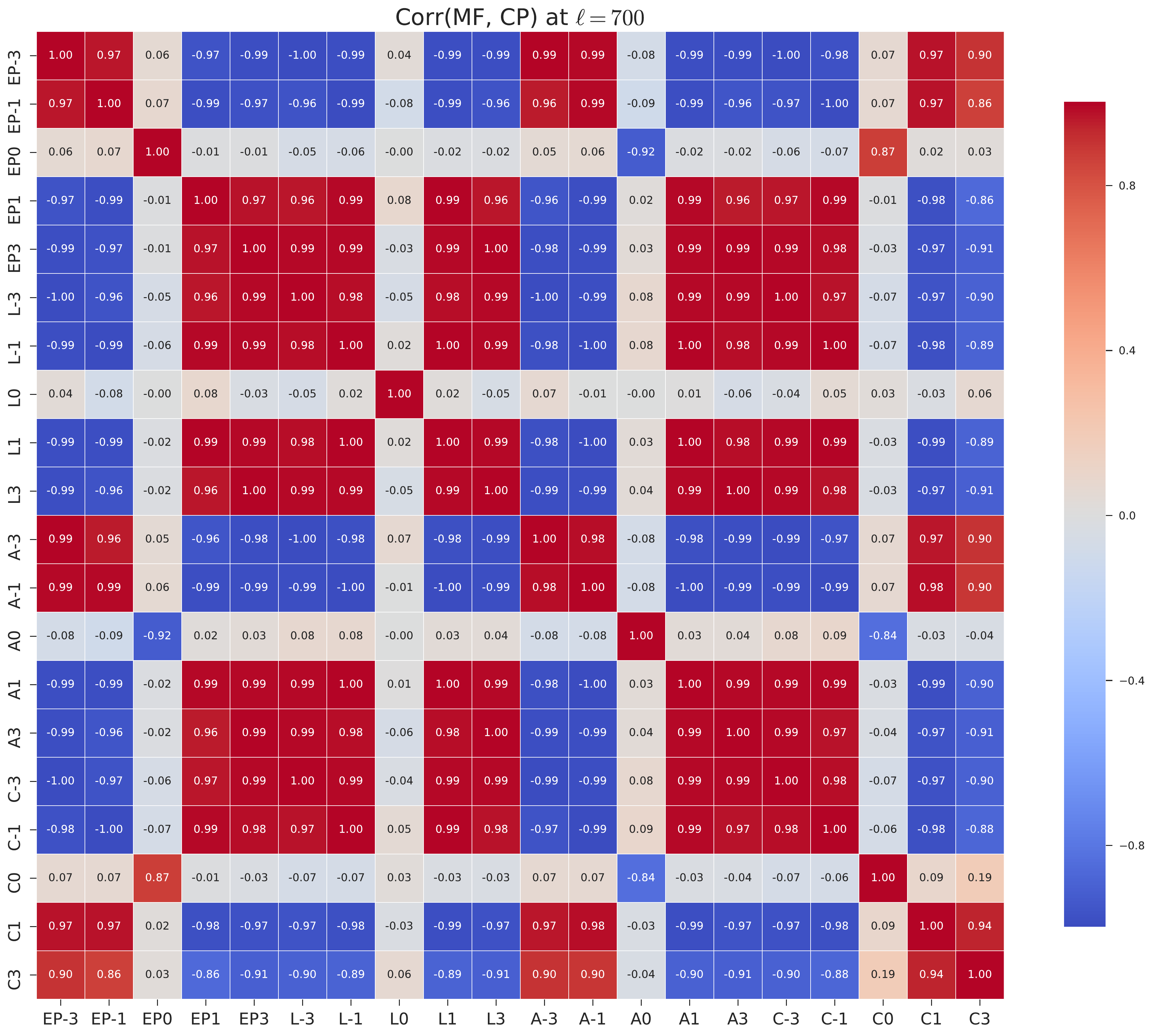}  
\caption{Correlation between and among the three Lipschitz-Killing curvatures and Critical points at different threshold values (as
  shown in the axis). Note the strong positive and negative correlation at $u \ne 0$. We wrote A$u$ for $u=3,1,0,-1,-3$ for the area functional evaluated at the level $u$, similarly L$u$ for the boundary length at level $u$ and EP$u$ for the Euler-Poincar\'e characteristic. The figure is realized setting $\ell=700$. \label{fig:corr}}
\end{center}
\end{figure}

\newpage
In figure \ref{fig:corr} 
we present our evidence on cross-correlations. As expected, correlations are very close to one (in absolute value) for any pair of random statistics evaluated at non-zero thresholds, including area, boundary length, Euler-Poincar\'e characteristic and the number of critical points; considering extrema (maxima and minima) and saddles separately would yield the same outcome. The simulations also confirm uncorrelation when expected, for instance between the nodal length (which is dominated by the fourth-order chaos, see \cite{MRW17}) and the defect, which is dominated by odd order chaoses (see \cite{MW14}). 

All these results have potential for applications in the statistical analysis of random fields, for instance when testing for nonGaussianity and isotropy or to search point-like sources/impurities in Cosmic Microwave Background radiation data. We do not address these issues in the present work, but we leave them as avenues for further research.

\section{Appendix}

In this Appendix, we report some numerical computations on constants needed for higher-order approximations on the behaviour of the limiting variances. 

Recall first that, denoting, as usual, $h_{\ell,q} :=\int_{\mathbb{S}^2} H_q(T_\ell(x)) \,dx$,
it is known (see for example \cite{MW14}), that for $q=3$ and $q\geq 5$, one has

$$\mbox{Var}(h_{\ell,q})=(4\pi)^2 q! \int_{0}^{\pi/2} P_\ell^q(\cos \theta )\sin \theta d\theta \sim (4\pi)^2 q! \frac{c_q}{\ell^2} $$
with
\begin{equation}\label{J0}
c_q=\int_{0}^{\infty} \psi J_0(\psi)^q d\psi \geq 0, \mbox{ } \mbox{ }\mbox{ } J_0(x)=\sum_{k=0}^{\infty} \dfrac{(-1)^kx^{2k}}{2^{2k}(k!)^2}
\end{equation} being the $J_0$ Bessel function. Moreover for $q=2,4,$ the order of magnitude of the corresponding variance is larger, namely: 

$$\mbox{Var}(h_{q;\ell}) \sim 16 \pi^2 \frac{1}{\ell} \mbox{  }\mbox{  }\mbox{ for } q=2$$
$$\mbox{Var}(h_{q;\ell}) \sim 576 \frac{\log \ell}{\ell^2}\mbox{  }\mbox{  } \mbox{ for } q=4.$$

Defining, as in (\ref{Cq}), $$C_q:=  \int_{0}^{L} J_0(\psi)^q \psi \,d\psi, \mbox{  } \mbox{ for } q=3 \mbox{ and }q\geq 5,$$
for the values $L=50$, 100, 200, we find, exploiting Matlab, the following numerical evaluations.\\

\begin{figure}[h!]
	\centering
	\includegraphics[width=0.8\linewidth]{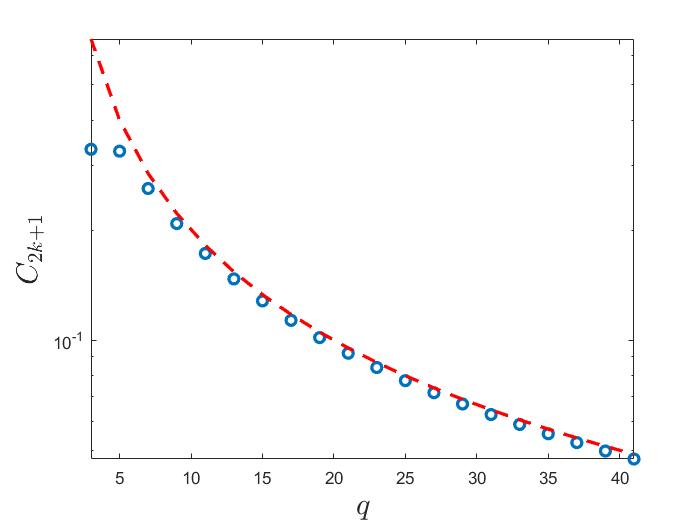}
	\caption{The red dashes represent the function $\frac{2}{q}$; whereas the blu circles, the coefficients $C_q$ for odd $q$. The plot is realized setting $L=100$.}
	\label{fig:cqcome2suqcopy}
\end{figure}

\begin{figure}[h!]
	\centering
	\includegraphics[width=0.8\linewidth]{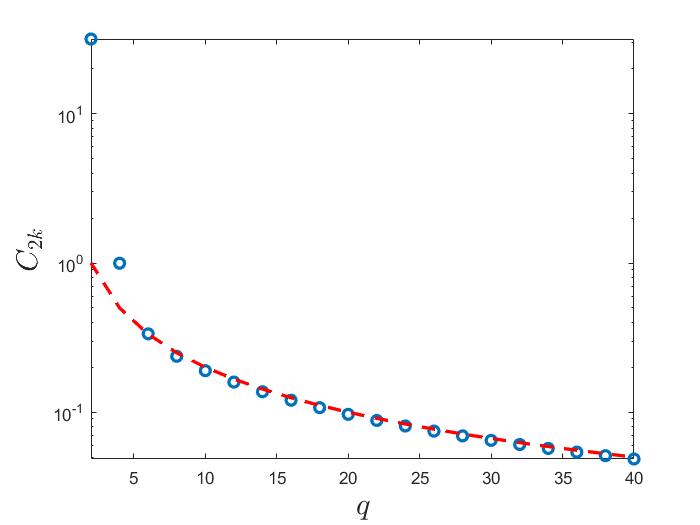}
	\caption{The red dashes represent the function $\frac{2}{q}$; whereas the blu circles, the coefficients $C_q$ for even $q$. The plot is realized setting $L=100$.}
	\label{fig:c2k}
\end{figure}

\begin{center}
\begin{tabular}{|c|c|c|c|}
\hline
	\textbf{ $C_q$} &\textbf{	$L=50$} & \textbf{$L=100$} & \textbf{$L=200$}\\
	\hline $C_5$ &	0.3286  & 0.3289 & 0.3290\\
	\hline $C_6$ &	0.3344  & 0.3352 & 0.3356\\
	\hline $C_7$ &	0.2600  & 0.2600 & 0.2600\\
	\hline $C_8$ &	0.2369 & 0.2369& 0.2369\\
	\hline $C_9$ &	0.2085  & 0.2085 & 0.2085\\
	\hline $C_{10}$ & 0.1897 & 0.1897&	0.1897\\
	\hline $C_{11}$ &	0.1727  & 0.1727 & 0.1727\\
	\hline $C_{12}$	& 0.1590 & 0.1590 & 0.1590 \\
	\hline $C_{13}$	& 0.1472 & 0.1472 & 0.1472 \\
	\hline $C_{17}$	& 0.1134 & 0.1134 & 0.1134 \\
	\hline $C_{18}$	& 0.1072 & 0.1072 & 0.1072 \\
	\hline $C_{24}$	& 0.0808 & 0.0808 & 0.0808 \\
	\hline $C_{25}$	& 0.0776 & 0.0776 & 0.0776 \\
	\hline
\end{tabular}
\end{center}
It can be seen from Figure \ref{fig:cqcome2suqcopy} and Figure \ref{fig:c2k} (realized for $L=100$) that the behavior of $C_q$, for $q\geq 5$ (odd or even), is well approximated by
\begin{equation}\label{Cq-approx}
C_q \sim \dfrac{2}{q}.
\end{equation}


\begin{figure}[h!]
	\centering
	\includegraphics[width=0.8\linewidth]{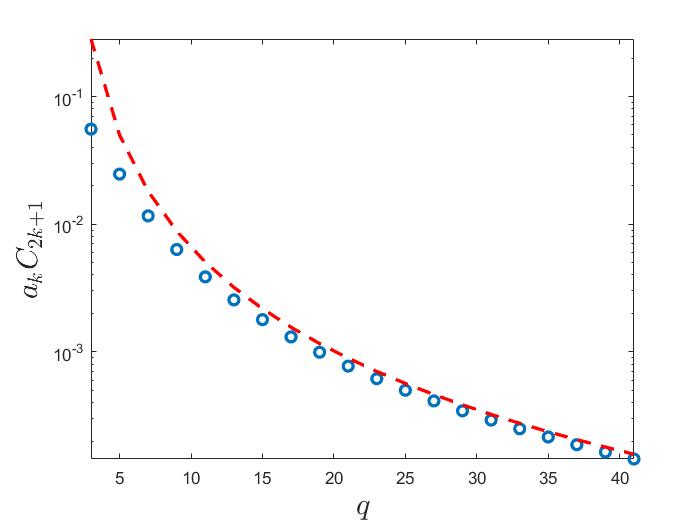}
	\caption{The red dashes represents the function $\dfrac{1}{2\sqrt{\pi}q^{5/2}}$; whereas the blue circles the points $a_kC_{2k+1}$. The plot is realized setting $L=100$.}
	\label{fig:akc2k1comeq5mezzi}
\end{figure}

\begin{remark}
	In \cite{MW12}, the asymptotic behavior of the coefficients $a_k$ is proved to be (using Stirling approximation) $\frac{a}{k^{3/2}}$, where the constant $a$ can be computed to be $\frac{1}{2\sqrt{\pi}}$; in view of (\ref{Cq-approx}) the product $a_k C_{2k+1}$ behaves as $\frac{1}{2\sqrt{\pi}k^{3/2}}\times \frac{2}{2k+1}$ and therefore as $\frac{1}{2\sqrt{\pi}k^{5/2}}$. Indeed, figure \ref{fig:akc2k1comeq5mezzi} compares the points $a_kC_{2k+1}$ with the function $\frac{1}{2\sqrt{\pi}q^{5/2}}$; the fit appears very good, for reasonably large values of $q$.
\end{remark}

Let us now try to improve the numerical approximation for the variance of the fourth-order chaos. We have shown numerically that
\begin{equation}\label{1}
    \lim_{\ell \rightarrow \infty} \big[\int_{0}^{L} J_0^4(\psi) \psi \, d\psi -\dfrac{3}{2\pi^2} \log \ell \big]=0.297
\end{equation}
Actually, to check the validity of this result we can estimate the difference semi-analytically. Indeed, splitting the domain of the integral in $[0,10]$ and $[10,L]$, we obtain 
\begin{equation}\label{diff}
\int_{0}^{\ell+1/2} J_0(\psi)^4 \psi \, d\psi -\dfrac{3}{2\pi^2} \log \ell= \int_{0}^{10} J_0(\psi)^4 \psi \, d\psi+\int_{10}^{L} J_0(\psi)^4 \psi \, d\psi -\dfrac{3}{2\pi^2} \log \ell.
\end{equation}
Now, exploiting the expansion of $J_0(x)$, as $x \rightarrow +\infty$ (see \cite{szego}), namely:
$$\sin(x+\frac{\pi}{4}) \sqrt{\frac{2}{\pi x}}- \cos(x+\frac{\pi}{4}) \frac{1}{x^{3/2}4 \sqrt{2\pi}}-\frac{9}{64 \sqrt{2\pi}} \frac{1}{x^{5/2}} \sin(x+\frac{\pi}{4})+ O\bigg(\frac{1}{x^{7/2}}\bigg)$$ and substituting it on the second term of (\ref{diff}), we get that the left-hand side in (\ref{diff}) is given by
$$\int_{0}^{10} J_0(\psi)^4 \psi \, d\psi + \int_{10}^{L} \Big[ \sin(x+\pi/4)\sqrt{\frac{2}{\pi x}}-\cos(x+\pi/4)\frac{1}{x^{3/2}\sqrt{2\pi}}+O(\frac{1}{x^{5/2}})\big]^4 x \,dx -\dfrac{3}{2\pi^2} \log \ell .$$ 
Expanding the fourth power we obtain
\begin{equation}\label{diff2}
\begin{split}
&\int_{0}^{10} J_0^4(\psi) \psi \,d\psi + \int_{10}^{L} \bigg( \frac{1-\cos (2x+\pi/2)}{2} \bigg)^2 \frac{4}{\pi^2 x^2} x \,dx + \\& - \int_{10}^{L} \frac{4\cos (x+\pi/4)}{x^{3/2}\sqrt{2\pi}} \sin(x+\pi/4)^3 \frac{2^{3/2}}{(\pi x)^{3/2}}x \,dx  +O(\frac{1}{x^2})-\frac{3}{2\pi^2} \log \ell,
\end{split}
\end{equation}
which is equal to
\begin{equation}
    \begin{split}
&\int_{0}^{10} J_0^4(\psi) \psi \,d\psi + \frac{1}{\pi^2} (\log L- \log 10)+ \int_{10}^{L} \frac{\cos^2 (2x+\pi/2)}{\pi^2 x} \,dx -\frac{2}{\pi^2} \int_{10}^{L} \frac{\cos(2x+\pi/2)}{x} \,dx \\&- \frac{8}{\pi^2} \int_{10}^{L} \frac{\cos (x+\pi/4)\sin(x+\pi/4)^3}{x^2} \, dx +O(\frac{1}{x^2})-\frac{3}{2\pi^2} \log \ell.
\end{split}
\end{equation}
Solving the integral of the square of the cosine, we get
\begin{equation}
    \begin{split}
 &\int_{0}^{10} J_0^4(\psi) \psi \,d\psi + \frac{1}{\pi^2} (\log L- \log 10)+ \frac{1}{2\pi^2} (\log L- \log 10)+ \frac{1}{2\pi^2} \int_{10}^{L} \frac{\cos (4x+\pi)}{x} \,dx -\\& \frac{2}{\pi^2} \int_{10}^{L} \frac{\cos(2x+\pi/2)}{x} \,dx- \frac{8}{\pi^2} \int_{10}^{L} \frac{\cos (x+\pi/4)\sin^3(x+\pi/4)}{x^2} \, dx +O(\frac{1}{x^2})-\frac{3}{2\pi^2} \log \ell 
  \end{split}
\end{equation}
and then the logarithm terms cancel, leading to the expression
\begin{equation}
    \begin{split}
= & \int_{0}^{10} J_0^4(\psi) \psi \,d\psi - \frac{3}{2\pi^2} \log 10 - \frac{1}{8 \pi^2} \bigg[ \frac{\sin(4x)}{x}\bigg]_{10}^{L}- \frac{1}{\pi^2} \bigg[\frac{\cos(2x)}{x} \bigg]_{10}^{L}- \frac{1}{8\pi^2} \int_{10}^L \frac{\sin(4x)}{x^2} \, dx \\& - \frac{1}{\pi^2} \int_{10}^{L} \frac{\cos(2x)}{x^2} \, dx +O(\frac{1}{x^2}).
\end{split}
\end{equation}
Now, using the fact that 
$$\int_{10}^{\infty} \frac{\sin(4x)}{x^2} \,dx = \frac{\sin 40}{10}- 4C_i(40)$$ and 
$$\int_{10}^{\infty} \frac{\cos 2x}{x^2} \,dx= 2S_i(20)-\pi +\frac{\cos 20}{10},$$ where 
$C_i(\cdot)$ and $S_i(\cdot)$ are the 
cosine and sine integral functions, respectively,
we can approximate (\ref{diff2}) with
\begin{equation}\label{diff3}
    \int_{0}^{10} J_0(\psi)^4 \psi \,d\psi - \frac{3}{2\pi^2} \log 10+ \frac{1}{2\pi^2} C_i(40)-\frac{2}{\pi^2} S_i(20)+ \frac{1}{\pi}
\end{equation}
and computing the value of $\int_{0}^{10} J_0(\psi)^4 \psi \,d\psi$ numerically, we find that (\ref{diff3}) is equal to 0.298, in very good agreement with the value in (\ref{1}).\\\\

Finally, we summarize, in the following tables, the analytic formulas used, for the Lipschitz-Killing curvatures, in the simulations.

\begin{table}[H]
\begin{center}
\begin{tabular}{|c|c|c|c}
	\hline
	\textbf{LKC } &\textbf{	Mean } & \textbf{ Variance }\\
	\hline $\mathcal{L}_{2}(A_{u}(f_{\ell };\mathbb{S}^{2}))$ & $1-\Phi(u)$& $\frac{1}{8\pi} u^2 e^{-u^2} \frac{1}{\ell} + O(\frac{\log \ell}{\ell^2})$ \\
	\hline $\mathcal{L}_{1}(A_{u}(f_{\ell };\mathbb{S}^{2}))$ & $\frac{1}{4\sqrt{2}} e^{-u^2/2} \sqrt{\ell(\ell+1)}$ & $\frac{1}{128} u^4 e^{-u^2} \ell+O(\log \ell)$ \\
	\hline $\mathcal{L}_{0}(A_{u}(f_{\ell };\mathbb{S}^{2}))$  & $ \frac{1}{4\pi}\sqrt{\frac{2}{\pi}} e^{-u^2/2}u \frac{\ell(\ell+1)}{2}+\frac{1}{2\pi}(1-\Phi(u))$ & $\frac{1}{128\pi^3}u^2(u^2-1)^2 e^{-u^2} \ell^3+O(\ell^2 \log \ell)$ \\
	\hline
\end{tabular}
	\caption{Theoretical expressions for the expected values and the variances of the three Lipschitz-Killing curvatures, Area ($\mathcal{L}_{2}$), Half of the Boundary Length Area ($\mathcal{L}_{1}$), Euler-Poincar\'e Characteristic ($\mathcal{L}_{0}$). Threshold level $u\ne 0$.} \label{table:lkcu}
\end{center}
\end{table}

\begin{table}[H]
\begin{center}
\begin{tabular}{|c|c|c|c}
	\hline
	\textbf{ LKC  } &\textbf{	Mean } & \textbf{ Variance }\\
	\hline $\mathcal{L}_{2}(A_{u=0}(f_{\ell };\mathbb{S}^{2}))$ & $\frac{1}{2} $& $  \frac{0.0188}{\ell^2}+o(\frac{1}{\ell^2})$  \\
	\hline $\mathcal{L}_{1}(A_{u=0}(f_{\ell };\mathbb{S}^{2}))$ & $\frac{1}{4\sqrt{2}} \sqrt{\ell(\ell+1)}$ & $\frac{1}{128} \frac{1}{16 \pi^2} \big\{ \log \ell+1.9542 \big\}+O(1)$ \\
	\hline $\mathcal{L}_{0}(A_{u=0}(f_{\ell };\mathbb{S}^{2}))$  & $\frac{1}{4\pi}$ & $O(\ell^2 \log \ell)$  \\
	\hline
	\end{tabular}
\caption{Theoretical expressions for the expected values and the variances of the three Lipschitz-Killing curvatures, Area ($\mathcal{L}_{2}$), Half of the Boundary Length ($\mathcal{L}_{1}$), Euler-Poincar\'e Characteristic ($\mathcal{L}_{0}$). Threshold level $u=0$.} \label{table:lkc0}
\end{center}
\end{table}

We stress again that here the area of the sphere has been normalized to $|\mathbb{S}^2|=1$, hence, to obtain these statistics when $|\mathbb{S}^2|=4\pi$ we need to multiply the mean for $4\pi$ and the variance for $16 \pi^2$, for the area and the Euler-Poincar\'e characteristic. For the boundary length we recall that there is a further factor 2 to take into account (boundary length $=2\mathcal{L}_{1}(A_{u}(f_{\ell };\mathbb{S}^{2}))$), so that we need to multiply for $4\pi \times 2$ to obtain the expected value and for $ 16\pi^2 \times 4$ for the variance. The asymptotic behavior at $u=0$ of the Euler Poincar\'e characteristic is easily seen to be $O(\ell^2 \log \ell)$ (exploiting results on extrema and saddles) but a rigorous evaluation of the leading constant is still missing.

We conclude reporting in the following table the formulae exploited for expected values and variances for the critical points, recalling that $$\Gamma(a,x)=\int_{x}^{\infty} t^{a-1}e^{-t}\,dt.$$
\begin{table}[H]
\begin{center}
\begin{tabular}{|c|c|c|c}
	\hline
	\textbf{ } &\textbf{	Mean } & \textbf{ Variance }\\
	\hline 
	$\mathcal{N}_{-\infty}^c(f_\ell)$ &$\frac{2}{\sqrt{3}} \ell^2+O(1)$ & $\frac{1}{3^3\pi^2} \ell^2 \log \ell +O(\ell^2)$ 
\\	\hline 
$\mathcal{N}_u^c(f_\ell)$& $\frac{2}{\sqrt{8\pi}} \ell(\ell+1) \big\{ \frac{2}{\sqrt{6}}\Gamma(\frac{1}{2},\frac{3u^2}{2})+e^{-u^2/2}u\big\}$ &
$\ell^3 \frac{1}{8\pi} e^{-3u^2} u^2 (2+e^{u^2} (u^2-1))^2 +O(\ell^2 \log \ell)$\\

	\hline $\mathcal{N}_{0}^c(f_\ell)$ & $\frac{1}{\sqrt{3}} \ell(\ell+1) +O(1)$ &  $\frac{1}{4 \pi^2 27} \ell^2 \log \ell+O(\ell^2)$\\
	\hline
\end{tabular}
	\caption{Theoretical expressions for the expected values and the variances of the critical points. Threshold level $u\ne 0$ and $u=0$.} \label{table:cp}
\end{center}
\end{table}

\section{Acknowledgments}
 We acknowledge the use of the National Energy Research Scientific Computing Center (NERSC) super-computing facilities. Maps and results
have been derived using the \healpix (http://healpix.jpl.nasa.gov)
software package developed by \cite{healpix}. Some of the theoretical results exploited in this paper were derived in collaboration with Igor Wigman, to whom we are grateful for many insightful discussions.


\bibliography{biblio1}


\end{document}